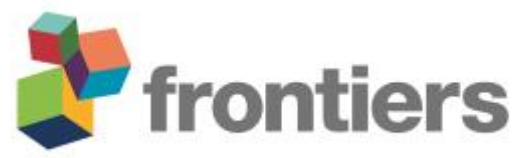

# Using PyBioNetFit to Leverage Qualitative and Quantitative Data in Biological Model Parameterization and Uncertainty Quantification

Ely F. Miller[1†], Abhishek Mallela[2,3†§], Jacob Neumann[4], Yen Ting Lin[2,5], William S. Hlavacek[2,3*], Richard G. Posner[1*]

[1]Department of Biological Sciences, Northern Arizona University, Flagstaff, AZ, United States of America

[2]Center for Nonlinear Studies, Los Alamos National Laboratory, Los Alamos, NM, United States of America

[3]Theoretical Biology and Biophysics Group, Theoretical Division, Los Alamos National Laboratory, Los Alamos, NM, United States of America

[4]Department of Chemistry and Chemical Biology, Cornell University, Ithaca, NY, United States of America

[5]Information Sciences Group, Computer, Computational and Statistical Sciences Division, Los Alamos National Laboratory, Los Alamos, NM, United States of America

[†] These authors contributed equally.

[§]Current address: Department of Mathematics, Dartmouth College, Hanover, NH, United States of America

* Address correspondence to W.S.H. (hlavacek@lanl.gov) or R.G.P. (richard.posner@nau.edu).

**Abstract**

Data generated in studies of cellular regulatory systems are often qualitative. For example, measurements of signaling readouts in the presence and absence of mutations may reveal a rank ordering of responses across conditions but not the precise extents of mutation-induced differences. Qualitative data are often ignored by mathematical modelers or are considered in an *ad hoc* manner, as in the study of Kocieniewski and Lipniacki (2013) [*Phys Biol* **10**: 035006], which was focused on the roles of MEK isoforms in ERK activation. In this earlier study, model parameter values were tuned manually to obtain consistency with a combination of qualitative and quantitative data. This approach is not reproducible, nor does it provide insights into parametric or prediction uncertainties. Here, starting from the same data and the same ordinary differential equation (ODE) model structure, we generate formalized statements of qualitative observations, making these observations more reusable, and we improve the model parameterization procedure by applying a systematic and automated approach enabled by the software package PyBioNetFit. We also demonstrate uncertainty quantification (UQ), which was absent in the original study. Our results show that PyBioNetFit enables qualitative data to be leveraged, together with quantitative data, in parameterization of systems biology models and facilitates UQ. These capabilities are important for reliable estimation of model parameters and model analyses in studies of cellular regulatory systems and reproducibility.

# 1 Introduction

In systems biology modeling, practitioners routinely disregard purely qualitative data—such as ordinal categorizations (e.g., low/medium/high), binary outcomes of case-control comparisons (e.g., up/down relative to a reference), and simple threshold crossing indicators (e.g., yes/no)—as well as semi-quantitative data, such as densitometric readouts of Western blots, which typically lack full quantification due to potential signal saturation and absence of calibration experiments. Bias against use of such data perhaps arises from the rich abundance of quantitative data (e.g., sequential measurements of state variables having high precision and fine time resolution) in fields such as physics, in which biological modelers are commonly trained. Nevertheless, qualitative/semi-quantitative data have been leveraged in influential biological modeling studies (Mitra and Hlavacek, 2019), such as those of Tyson and co-workers focused on modeling yeast cell-cycle control (Chen et al., 2000; Chen et al., 2004; Kraikivski et al., 2015; Barik et al., 2016; Tyson et al., 2019). In these studies, qualitative/narrative descriptions of yeast mutant phenotypes were used to constrain parameter estimates. Other related studies include those of Kinney and co-workers focused on using noisy semi-quantitative readouts of massively parallel reporter assays to inform the modeling of regulation of gene expression (Kinney et al., 2007; 2010; Atwal and Kinney, 2016; Kinney and McCandlish, 2019).

In recent years, there has been increasing appreciation of the need to leverage all types of available data in biological modeling, as evidenced by the development and/or application of various approaches for using qualitative/semi-quantitative data in model parameterization, optionally in combination with quantitative data. Demonstrated approaches include likelihood-free inference conditioned on binary data (Toni et al., 2011), data-transformation methods (e.g., optimal scaling) (Pargett and Umulis, 2013; Pargett et al., 2014; Schmiester et al., 2020; 2021; Dorešić et al., 2024), information-theoretic approaches (Atwal and Kinney, 2016; Tareen et al., 2022), constrained heuristic optimization in which qualitative observations are formalized as constraints (Oguz et al., 2013; Mitra et al., 2018), and (multinomial) probit regression (Mitra and Hlavacek, 2020). Software tools developed to cope with qualitative/semi-quantitative data include PyBioNetFit (Mitra et al., 2019; Mitra and Hlavacek, 2020), pyPESTO (Schmiester et al., 2021; Schälte et al., 2023; Dorešić et al., 2024), and MAVE-NN (Tareen et al., 2022).

Here, in a new demonstration of the ability of PyBioNetFit to leverage both quantitative and qualitative data, we redo the parameterization of an ordinary differential equation (ODE) model developed by Kocieniewski and Lipniacki (2013) to study the roles of MEK isoforms in ERK activation. Originally, this model was parameterized by tedious trial-and-error manual tuning of parameter values for consistency with a mix of published qualitative and quantitative data (Catalanotti et al., 2009; Kamioka et al., 2010). The qualitative data included orderings of low-resolution readouts measured for parental and perturbed/mutant cell lines. In our re-analyses of these data, we define qualitative data unambiguously using the Biological Property Specification Language (BPSL) (Mitra et al., 2019), which increases the reusability of these data, and we leverage both the qualitative and quantitative data using various features of PyBioNetFit, which yields a systematic and automated approach to parameter estimation and uncertainty quantification (UQ). This approach increases reproducibility. Furthermore, we clarify that the parameter estimation method of Mitra et al. (2018), which is implemented in PyBioNetFit (Mitra et al., 2019) and which we leveraged in our study, can be viewed as a likelihood-based inference procedure. Through metaheuristic optimization, we obtain maximum likelihood estimates for parameter values, and for UQ, we perform profile likelihood analysis (Kreutz et al., 2013). In addition, through application of an adaptive Markov chain Monte Carlo (MCMC) sampling algorithm (Andrieu and Thoms, 2008) implemented in PyBioNetFit (Neumann et al., 2022), we obtain samples representing a Bayesian parameter posterior, which allows us to quantify both parametric and prediction uncertainties. Prediction uncertainties are quantified by posterior predictive distributions. Thus, we demonstrate UQ in two ways, each of which considers both quantitative and qualitative data. The absence of UQ was a notable limitation of the original study of Kocieniewski and Lipniacki (2013), a limitation rooted in the *ad hoc* labor-intensive parameterization approach of that study.

The results presented here provide a new illustration of how PyBioNetFit facilitates efficient and reproducible use of qualitative data (together with quantitative data) in parameterization of systems biology models. The results also provide an illustration of PyBioNetFit capabilities useful for UQ, which is essential for a variety of reasons, from assessment of model credibility to model improvement. This report should be useful to researchers who wish to leverage qualitative data in parameterization of biological models. For the convenience of readers unfamiliar with specialized terminology (e.g., BNGL, BPSL, UQ), we provide a glossary of key terms in the Supplementary Material.

## 2 Materials and Methods

### 2.1 Model for parental cell line and model variants for mutant cell lines

Kocieniewski and Lipniacki (2013) defined their model for normal, or wild type (WT), MEK1 signaling using the conventions of the BioNetGen language (BNGL) (Faeder et al., 2009) and made the model available online in the form of a BNGL file (a plain-text file with a .bngl filename extension) within a ZIP archive (https://iopscience.iop.org/article/10.1088/1478-3975/10/3/035006/data). Annotation at the top of this BNGL file describes how to modify the model to account for MEK1 knockout (KO) and three MEK1 mutations: N78G, which ablates MEK1 dimerization with itself and MEK2; T292A, which ablates ERK-mediated inhibitory phosphorylation of MEK1 and concomitant negative feedback; and T292D, a phosphomimetic mutation. From this starting point, we created separate BNGL files for parental and mutant cell lines labeled WT, N78G, T292A, and T292D. The resulting BNGL files, MEK1_WT.bngl, MEK1_KO.bngl, MEK1_N78G.bngl, MEK1_T292A.bngl, and MEK1_T292D.bngl, are freely available online (https://github.com/lanl/PyBNF/tree/master/examples/Miller2025_MEK_Isoforms) and can be processed by PyBioNetFit (Mitra et al., 2019), version 1.1.9. In PyBioNetFit workflows, models can be defined using either BNGL or Systems Biology Markup Language (SBML) (Keating et al., 2020). BioNetGen (Harris et al., 2016) can convert a BNGL file to an SBML file. SBML files are plain-text files with .xml filename extensions.

### 2.2 Data used in model parameterization

Kocieniewski and Lipniacki (2013) carefully identified the data that they used in model parameterization. We collected these same data from the primary sources, namely the reports of Catalanotti et al. (2009) and Kamioka et al. (2010). Each qualitative observation from the case-control comparisons of Catanotti et al. (2009)—e.g., up/down relative to a reference at a particular time after initiation of signaling—was formalized using the Biological Property Specification Language (BPSL) (Mitra et al., 2019). BPSL statements, which can be interpreted by PyBioNetFit, were then collected in PROP files (plain-text files with .prop filename extensions). See BPSL examples and breakdowns in the box below, and all BPSL statements used in our study are available in the supplementary tables 1-5. Quantitative observations, time-series data, from the study of Kamioka et al. (2010) were collected/tabulated in an EXP file (a plain-text file with a .exp filename extension). It should be noted that, in an EXP file, "nan" indicates a missing measurement. The resulting EXP and PROP files are freely available online (https://github.com/lanl/PyBNF/tree/master/examples/Miller2025_MEK_Isoforms) and can be processed by PyBioNetFit (Mitra et al., 2019), version 1.1.9. It should be noted that each EXP and PROP file is meant to be used with a particular model variant, as indicated by the labels WT, N78G, T292A, and T292D. For example, during model parameterization, the data collected in the WT-labeled EXP and PROP files (WT.exp and WT.prop) are compared against the corresponding outputs of the WT model (defined in the MEK1_WT.bngl file). The relevant model outputs are identified in accordance with established conventions (Mitra et al., 2019); for example, in an EXP file, times are indicated by row labels and each column header has the name of an observable or function defined in a BNGL file. We note that the only non-empty EXP file maps to the WT model, because quantitative, time-series data was only available for normal signaling (i.e., signaling unperturbed by MEK1 knockout or mutation).

| BPSL Statement | Prose Statement |
|---|---|
| $\text{WT. MEK_pRDS at time} = 300 < \text{N78G. MEK_pRDS at time} = 300$ | The copy number of RAF-phosphorylated (pRDS) normal/wildtype (WT) MEK (in parental cells) is less than the copy number of RAF-phosphorylated MEK with mutation N78G (in a derived cell line) at 300 s after stimulation of RAS/RAF/MEK/ERK signaling. |
| $\text{N78G. pERK1_2_wt at time} = 300 > \text{T292D. pERK1_2_wt at time} = 300$ | The copy number of MEK1/2-double phosphorylated ERK with mutation N78G at 300 seconds is less than MEK-double phosphorylated ERK with mutation T292D at 300 seconds after simulation of RAS/RAF/MEK/ERK signaling. |

### 2.3 Maximum likelihood estimation and likelihood profiling

Across the five model variants under consideration (WT, KO, N78G, T292A, and T292D), there are 46 total parameters. We took 28 of these to be adjustable and set the other 18 at fixed values specified by Kocieniewski and Lipniacki (2013). Six of these parameters ($s_1, d_1, s_2, d_2, c_1, c1_{init}$) correspond to the simulating the initial concentrations of EGFR, Sos1, and EGFR/ligand subcomplexes. In preliminary analyses, we found that estimating these parameters directly added unnecessary computational complexity without altering model behavior. Instead, we determined their peak concentrations from early simulation runs and fixed those values before initiating the cascade. This approach simplified parameterization while preserving biological realism. The remaining 12 fixed parameters ($X, MEK_Fraction, EGFR_0, SOS1_0, RAS_0, RAF_0, MEK_{tot_0}, MEK1_0, MEK2_0,$ $MEK1_T292p_0, ERK_0, PHP_MEK_0$) define the initial conditions of the resting-state cell and mutants. These were held at the original values from Kocieniewski and Lipniacki (2013) to maintain comparability with their published results.

Furthermore, all 28 adjustable parameters chosen to be used in our study were initially centered on Kocieniewski and Lipniacki's (2013) parameter values and given bounds of one or two orders of magnitude (10x - 100x) in either direction of the initial value. Some parameter bounds were further adjusted if the algorithm began pushing the limit of a given bound. To ensure biological meaningfulness, we imposed physical constraints where applicable (e.g., rate constants bounded below by zero and above by the diffusion limit; fractional parameters bounded between 0 and 1). We made a conscious effort making these constraint choices to help maintain realism while supporting numerical stability in both optimization and Bayesian inference. All parameter bounds can be seen in Table 2. In addition, we introduced 3 adjustable scaling factors, which are useful for comparing WT model outputs to serial measurements. Consequently, in parameter estimation, we considered a total of 31 adjustable parameters, which we will denote as $\theta$. To obtain point estimates, $\hat{\theta}$, for the 31 adjustable parameters $\theta$, we minimized an objective function $F(\theta) = F_{\text{qual}}(\theta) + F_{\text{quant}}(\theta)$, the form of which was previously described by Mitra et al. (2018). The objective function accounts for all qualitative data (a total of 90 BPSL statements) and all quantitative data (a total of 18 serial measurements). We constrained $\hat{\theta}$ to lie within a feasible region of parameter space, which we will denote as $\Theta$. The feasible region $\Theta$ was defined by box constraints (lower and upper bounds) applied to each adjustable parameter. In summary, we found

$$\hat{\theta} = \arg\min_{\theta \in \Theta} F(\theta) \quad (1)$$

Thus, $\hat{\theta}$ is the product of a global fit. As explained below, the objective function $F(\theta)$ is related to a likelihood, and moreover, minimizing the objective function maximizes this likelihood. Thus, our point estimates are maximum likelihood estimates (MLEs).

The fitting problem setup, including identification of each adjustable parameter and the corresponding box constraints, and workflow were defined with a PyBioNetFit configuration, or CONF, file. This file (a plain-text file with a .conf filename extension) is freely available online (https://github.com/lanl/PyBNF/tree/master/examples/Miller2025_MEK_Isoforms/MEK_isoform_optimization_DE). In parameter estimation, the objective function $F(\theta)$ was minimized using PyBioNetFit's implementation of a parallelized differential evolution (DE) algorithm (Mitra et al., 2019). PyBioNetFit-enabled fitting runs were executed within an institutional high-performance computing environment, on the Monsoon cluster at Northern Arizona University. We used 25 of 28

CPUs on a single node within the cluster and the average wall-clock time for optimization simulations were approximately 2 minutes and conducted 2,825 objective function evaluations. The model of CPUs used are Intel(R) Xeon(R) CPU E5-2680 v4 CPUs. Multiple runs were performed, with each run starting at a randomly chosen point within the feasible region of parameter space $\Theta$. The CONF file includes algorithmic parameter settings (e.g., population size) and maps EXP and PROP files to model variants.

Profile likelihood analysis (Kreutz et al., 2013) was performed as described by Mitra et al. (2018). Computation of a profile involves repeatedly solving the parameterization problem described above, but with one selected parameter in $\theta$ held fixed at a specified value, which is varied across the profile. This analysis is available in the supplemental material as supplemental figure 4. We found that 14 of the 28 parameters selected for maximum likelihood estimation were identifiable, while the remaining parameters showed little sensitivity to perturbation from their original values.

When included in maximum likelihood estimation (MLE), unidentifiable parameters can reduce the identifiability of sensitive parameters if they are not fixed during parameterization. To assess whether estimating all 28 parameters affected PyBioNetFit's parameterization of the MEK isoform models, we performed two fits. In the first, MLEs were estimated for all 28 parameters. In the second, only the 14 identifiable parameters were estimated, while the remaining were fixed to the original values from Kocieniewski and Lipniacki (2013). Using PyBioNetFit's differential evolution algorithm and sum of squares objective function, we found that fixing the non-sensitive parameters did not significantly alter the overall objective score or the estimates of identifiable parameters.

### 2.4 Bayesian inference and uncertainty quantification

In Bayesian inference and uncertainty quantification, for simplicity, we considered only two adjustable model parameters: $d_3$, a rate constant characterizing degradation of ligand-bound epidermal growth factor receptor (EGFR) dimers, and $u_3$, a rate constant characterizing phosphatase activity responsible for reversal of ERK-mediated negative-feedback phosphorylation of SOS1. Both parameters were deemed practically identifiable by profile likelihood. The adjustable parameters also included three scaling factors, as in fitting. Finally, the adjustable parameters included a noise model parameter, $\sigma$. In Bayesian inference, we used a likelihood function described by Mitra and Hlavacek (2020), which has $\sigma$ as a hyperparameter, and a proper uniform prior defined by box constraints on each of the six adjustable parameters (see below).

Bayesian inference was enabled by the adaptive Markov chain Monte Carlo (MCMC) sampler implemented in PyBioNetFit (Neumann et al., 2022), version 1.1.9. The sampling problem setup and workflow were defined with a PyBioNetFit configuration, or CONF file. This file (a plain-text file with a .conf filename extension) is freely available online (https://github.com/lanl/PyBNF/tree/master/examples/Miller2025_MEK_Isoforms/MEK_isoform_aMCMC). The CONF file specifies the box constraints that define the prior. Each sampling job included 25,000 burn-in iterations, 25,000 adaptation iterations (used to tune the covariance matrix of the proposal kernel), and 250,000 production iterations. We generated five independent chains. Each chain was initialized at a random point within the prior. All five chains converged to the same posterior distribution. Convergence was evaluated by inspecting diagnostic trace plots and pairs plots. We also calculated convergence metrics (Vehtari et al., 2021) using the rstan R package. These metrics indicated convergence according to the guidance of Vehtari et al. (2021). PyBioNetFit-enabled sampling jobs were executed within an institutional high-performance computing environment, on the Monsoon cluster at Northern Arizona University. In the Monsoon cluster at Northern Arizona University, we used 25 CPUs of a single 28 CPU node containing Intel(R) Xeon(R) CPU E5-2680 v4 CPUs, and the approximate wall-clock time for our Bayesian inference simulations were 5.15 days, having conducted 1.25 million objective function evaluations.

### 2.5 Derivation of likelihood function

In this study, we consider a set of model variants $\mathcal{M}$ and a dataset $\mathcal{D} = \{y, z\}$. The dataset consists of $n$ quantitative observations, $y = \{y_1, \dots, y_n\}$, and $m$ qualitative observations, $z = \{z_{n+1}, \dots, z_{n+m}\}$. As a simplification, we assume that the $n + m$ observations are independent.

The quantitative observations $y$ are relative intensity measurements, which have been scaled such that $y_i \in (0,1]$ for $i = 1, \dots, n$. There are no replicate measurements, and each $y_i$ corresponds to a unique measurement condition, $c_i$. For each experimental readout $y_i$, there is a corresponding model output $f(c_i, \theta)$, which depends on condition $c_i$ and $P$ adjustable parameters, $\theta = \{\theta_1, \dots, \theta_P\}$. The model output $f(c_i, \theta)$ is generated by the model variant in $\mathcal{M}$ matched to condition $c_i$, which is always the WT model (because time-series data are not available for any of the mutant cells).

Each qualitative observation $z_i \in \{0,1\}$ $(i = n + 1, \dots, n + m)$ is the binary outcome of a comparison of two semi-quantitative measurements $\mathcal{A}_i$ and $\mathcal{B}_i$ made for two different cell lines (e.g., parental and mutant cells) or at two different time points within the same cellular background. By convention, we take $z_i = 0$ to indicate $\mathcal{A}_i < \mathcal{B}_i$ and $z_i = 1$ to indicate $\mathcal{A}_i \geq \mathcal{B}_i$. The measurements $\mathcal{A}_i$ and $\mathcal{B}_i$ are made at conditions $a_i$ and $b_i$, which are identical except for the difference in cell type or time. For each pair of measurements $(\mathcal{A}_i, \mathcal{B}_i)$, there is a pair of model outputs $(g(a_i, \theta), g(b_i, \theta))$. The output $g(a_i, \theta)$ is generated by the model variant in $\mathcal{M}$ matched to condition $a_i$, which encompasses cell type or time, and similarly, the output $g(b_i, \theta)$ is generated by the model variant matched to $b_i$. Moreover, for each $z_i$, we have a corresponding model prediction $H(g(a_i, \theta) - g(b_i, \theta))$, where $H$ is the Heaviside function.

In maximum likelihood estimation, we want to find $\hat{\theta} = \text{argmax}_{\theta \in \Theta} \mathcal{L}(\theta|\mathcal{D})$, where the likelihood $\mathcal{L}(\theta|\mathcal{D}) = P(\mathcal{D}|\theta)$ is the probability density over the data $\mathcal{D}$ given model structure and parameter settings. In inference, we view $\mathcal{L}(\theta|\mathcal{D})$ as a function of the adjustable parameters $\theta$ and take $\mathcal{L}(\theta|\mathcal{D})$ to express how probable the data are for given parameter settings. If observations are independent, $\mathcal{L}(\theta|\mathcal{D}) = \mathcal{L}(\theta|\{y, z\})$ decomposes into the product $\mathcal{L}(\theta|y)\mathcal{L}(\theta|z)$, where $\mathcal{L}(\theta|y)$ is the likelihood of the quantitative data and $\mathcal{L}(\theta|z)$ is the likelihood of the qualitative data.

Let us use $Y_i$ to denote a continuous random variable representing the distribution of possible quantitative measurement outcomes of an experiment performed at condition $c_i$. We take the observation $y_i$ to be a realization of $Y_i$, and we take the model output $f(c_i, \theta)$ to be a prediction of the expected measurement outcome at condition $c_i$. Formally, we take $f(c_i, \theta)$ to model $E[Y_i]$. Let us assume normally distributed measurement noise: $Y_i \sim N(f(c_i, \theta), \sigma_i^2)$. It then follows that

$$P\big(Y_i = y_i \big| f(c_i, \theta)\big) = \frac{1}{\sigma_i \sqrt{2\pi}} \exp\left[-\frac{1}{2}\left(\frac{y_i - f(c_i, \theta)}{\sigma_i}\right)^2\right] \tag{2}$$

If the quantitative data $y = \{y_i\}_{i=1}^{n}$ are taken to be independent, we find

$$P(y|\{f(c_i,\theta)\}_{i=1}^{n}) = \prod_{i=1}^{n} \frac{1}{\sigma_i\sqrt{2\pi}} \exp\left[-\frac{1}{2}\left(\frac{y_i - f(c_i,\theta)}{\sigma_i}\right)^2\right] \tag{3}$$

We can identify $P(y|\{f(c_i,\theta)\}_{i=1}^{n})$ as $\mathcal{L}(\theta|y)$. In the absence of information about variances, let us assume that $\sigma_i^2 = \sigma^2 > 0$ for $i = 1, \ldots, n$. Under this assumption of homoscedasticity (i.e., equal variances), we find

$$-\ln \mathcal{L}(\theta|y) = \frac{n}{2}\ln(2\pi\sigma^2) + \frac{1}{2\sigma^2}\sum_{i=1}^{n}\left(y_i - f(c_i,\theta)\right)^2 \tag{4}$$

Furthermore,

$$-\ln \mathcal{L}(\theta|y) \propto F_{\text{quant}} \equiv \sum_{i=1}^{n}\left(y_i - f(c_i,\theta)\right)^2 \tag{5}$$

Note that $\mathcal{L}(\theta|y)$ is maximized if we minimize $F_{\text{quant}}$. The expression for $F_{\text{quant}}$ is equivalent to Equation (2) in Mitra et al. (2018).

Let us use $Z_i, i \in \{n+1, \ldots, m+n\}$, to denote a discrete random variable representing the distribution of possible outcomes from a comparison of two semi-quantitative experimental readouts $\mathcal{A}_i$ and $\mathcal{B}_i$ made at conditions $a_i$ and $b_i$. There are only two possible outcomes, which are mutually exclusive: $Z_i = 0$ (indicating that $\mathcal{A}_i < \mathcal{B}_i$) or $Z_i = 1$ (indicating that $\mathcal{A}_i \geq \mathcal{B}_i$). We take $z_i$ to be a realization of $Z_i$. We will assume the following noise model: $Z_i \sim \text{Bernoulli}(p_i)$. Equivalently,

$$\Pr(Z_i = 0) = 1 - p_i \;\text{ and }\; \Pr(Z_i = 1) = p_i \tag{6}$$

Note that $p_i$ is the expectation $E[Z_i]$. Recall that the model prediction of $z_i$ is $H(g(a_i,\theta) - g(b_i,\theta))$. As in standard logistic regression, we will assume that the parameter $p_i$ is a sigmoid function of the difference $\delta_i \equiv g(a_i,\theta) - g(b_i,\theta)$:

$$p_i = \frac{1}{1 + \exp[-\delta_i/s_i]} \tag{7}$$

where $s_i > 0$ is a scale parameter. With this approach, we are assuming that the difference $\delta_i$ explains the probability $p_i$. If $\delta_i \geq 0$, then $p_i$ lies between 0.5 and 1 and $z_i = 1$ is more likely than $z_i = 0$. Conversely, if $\delta_i < 0$, then $p_i$ lies between 0 and 0.5 and $z_i = 1$ is less likely than $z_i = 0$. From the above considerations, we find

$$P(Z_i = z_i|\delta_i, s_i) = \left(\frac{1}{1 + e^{-\delta_i/s_i}}\right)^{z_i}\left(\frac{e^{-\delta_i/s_i}}{1 + e^{-\delta_i/s_i}}\right)^{1-z_i} \tag{8}$$

If the data $z = \{z_i\}_{i=n+1}^{n+m}$ are independent, we find

$$P(z|\{\delta_i\}_{i=n+1}^{n+m}, \{s_i\}_{i=n+1}^{n+m}) = \prod_{i=n+1}^{n+m}\left(\frac{1}{1 + e^{-\delta_i/s_i}}\right)^{z_i}\left(\frac{e^{-\delta_i/s_i}}{1 + e^{-\delta_i/s_i}}\right)^{1-z_i} \tag{9}$$

We can identify $P(z|\{\delta_i\}_{i=n+1}^{n+m}, \{s_i\}_{i=n+1}^{n+m})$ as $\mathcal{L}(\theta|z)$. After simplifications, we find

$$-\ln \mathcal{L}(\theta|z) = \sum_{i=n+1}^{n+m} \left[\ln\left(1 + e^{-\delta_i/s_i}\right) + (1 - z_i)\delta_i/s_i\right] \tag{10}$$

Furthermore, under an assumption that $\left|\frac{\delta_i}{s_i}\right| \gg 1$ for $i = n + 1, \ldots, n + m$, we find

$$-\ln \mathcal{L}(\theta|z) \approx F_{\text{qual}} \equiv \sum_{i=n+1}^{n+m} w_i[\max(0, -\delta_i) + (1 - z_i)\delta_i] \tag{11}$$

where $w_i = 1/s_i$. The $i$th term in the sum can be viewed as a static penalty function with weight $w_i$. The penalty is $w_i \cdot |\delta_i|$ if the explanatory variable $\delta_i$ is inconsistent with the observation $z_i$ and 0 otherwise. In the absence of information indicating how categorizations vary over a range of values for the explanatory variable, the weights can be set heuristically as described by Mitra et al. (2018). The above expression for $F_{\text{qual}}$ is equivalent to Equation (3) in Mitra et al. (2018).

In Equation (1), $F(\theta)$ is equal to $F_{\text{quant}}(\theta) + F_{\text{qual}}(\theta)$, where $F_{\text{quant}}(\theta)$ is given by Equation (5) and $F_{\text{qual}}(\theta)$ is given by Equation (11).

## 3 Results

Kocieniewski and Lipniacki (2013) developed a collection of related ordinary differential equation (ODE) models for ERK activation dynamics. These models capture and explain MEK isoform-specific effects in cell lines in which MEK1 is normally expressed (WT), knocked out (KO), and mutated at single amino-acid residues (N78G, T292A, and T292D). The models reproduce experimentally characterized signaling behavior, including various qualitative system properties; however, their parameter values were determined through an *ad hoc* manual trial-and-error procedure, precluding straightforward reuse of the qualitative data (up/down assays relative to a reference), reproducibility of the model parameterization approach, and any formal assessment of parametric and prediction uncertainty.

To demonstrate a better approach to model parameterization, we formalized qualitative system properties as Biological Property Specification Language (BPSL) statements (Mitra et al., 2019) (Supplemental Tables 1–5). Then, following the inference approach of Mitra et al. (2018), which is elaborated above, we applied, in a single global optimization, a parallelized metaheuristic optimization method implemented in PyBioNetFit (Mitra et al., 2019) to find maximum likelihood estimates (MLEs) for 28 model parameters and 3 scaling factors that relate model outputs to relative measurements (Table 2). The quality of fit to quantitative time-series data (relative measurements) from Kamioka et al., (2010) is illustrated in Figure 1. The quality of fit to qualitative data—up/down assays relative to a reference—from Catalanotti et al., (2009) is illustrated in Figures 2 and 3.

In Figure 1, the left panel (Figure 1A) shows the original fit of Kocieniewski and Lipniacki (2013) to time-series data. The right panel (Figure 1B) shows the comparable fit found in this study. Our automated parameterization achieves comparable fits across all time courses, with slightly improved agreement for SOS1 and ERK phosphorylation. This improvement is supported quantitatively by the RMSE values reported in Supplementary Tables S8 and S9, which show overall lower errors for our MLE parameter estimates relative to the original values from Kocieniewski and Lipniacki (2013). Note that the model outputs underlying the relative quantities plotted in Figure 1 are the absolute cellular abundances of phosphorylated EGFR, SOS1, and ERK. These quantities are each multiplied by an adjustable scaling factor to align with the relative measurements reported by Kamioka et al. (2010). Finally, objective function score outputs from PyBioNetFit's sum of squares (sos) objective function are available in supplemental table S7. PyBioNetFit's (figure 1B) final objective function score during optimization is 24.0, where Kocieniewski and Lipniacki's (2013) final objective function score is 40.0.

The curves in Figures 2 and 3 show calibrated model-predicted phosphorylated MEK and ERK, respectively, as a function of time in different cell lines, as indicated by color and pattern. At the time points corresponding to dotted vertical lines, readouts in pairs of distinct cell lines were compared, resulting in up or down scoring (Catalanotti et al., 2009). The empirical outcomes of these comparisons are indicated graphically above the dotted vertical lines. As in Figure 1, the panels at left (panels A, C, E, and G in both Figures 2 and 3) show the simulation results of Kocieniewski and Lipniacki (2013), and the panels at right (panels B, D, F, and H in both Figures 2 and 3) show the simulation results of this study. As can be seen, we obtained comparable consistency with empirical up/down scoring. Indeed, the calibrated model-predicted time-courses from this study and the earlier study are remarkably similar, despite differences in parameter estimates. To further show that PyBioNetFit's parameterization of the MEK isoform models maintained comparable consistency to Kocieniewski and Lipniacki's (2013) parameterization of the models, both the original parameterization and PyBioNetFit's parameterization satisfied 84/90 BPSL constraints determined from distinct cell line readouts from Catalanotti et al., (2009).

We note that Figures 2 and 3 only consider comparisons of readouts in different cell lines. Additional qualitative data were considered in fitting. These data were generated from comparisons of readouts at different time points within the same cellular background. Supplementary Tables 1–5 provide a full listing of the qualitative observations, formalized as BPSL statements, used in fitting.

Given the discrepancy between the parameter estimates found here and those of the original study, questions of parametric uncertainty naturally arise. To illustrate that our approach allows for uncertainty quantification (UQ), we generated profile likelihood plots for two rate constants in the models, $d_3$ and $u_3$ (Figure 4). These plots show that both parameters are practically identifiable and that estimation of the value for $d_3$ is more constrained by the available data than is estimation of the value for $u_3$. In other words, the width of any horizontal line segment interior to the profile of Figure 4A (solid blue curve) is shorter than the corresponding line segment in Figure 4B. Note that the dotted red vertical line in each panel marks the best-fit estimate.

To further demonstrate the capabilities of PyBioNetFit, we executed a Bayesian approach to inference and UQ, which leveraged the same datasets as those considered in maximum likelihood estimation. However, for the sake of simplicity, we focused on a smaller set of adjustable parameters. Through Markov chain Monte Carlo (MCMC) sampling, we were able to obtain samples converging on and representing the parameter posterior distribution. Sampling convergence metrics are reported in Table 3. Additional convergence diagnostics, likelihood and parameter trace plots and pairs plots, are shown in Supplementary Figures 1–3. Marginal posterior distributions for the rate constants $d_3$ and $u_3$ are shown in Figure 5. By propagating parametric uncertainty through simulations, we obtained posterior predictive distributions characterizing uncertainty in model predictions (Figures 6 and 7). Supplementary Table 6 quantifies consistency with qualitative observations.

# 4 Discussion

In this study, we demonstrated that PyBioNetFit (Mitra et al., 2019) can replace manual trial-and-error parameter tuning, as in the study of Kocieniewski and Lipniacki (2013), with a reproducible, automated pipeline that integrates both quantitative and qualitative data. We showed how to formalize up/down results of case-control comparisons as Biological Property Specification Language (BPSL) statements, and we showed how to leverage BPSL statements in concert with conventional time-series data in not only model parameterization but also rigorous uncertainty quantification (UQ). We elaborated on the approach of Mitra et al. (2018), showing that this approach is grounded in likelihood-based inference. We also demonstrated both frequentist and Bayesian approaches to rigorous uncertainty quantification (UQ), which was missing in the original modeling study of Kocieniewski and Lipniacki (2013).

An important aspect of our re-parameterization of Kocieniewski and Lipniacki's (2013) MEK isoform models is that different parameter sets can yield comparably good fits to the available data. This phenomenon arises when multiple combinations of parameters produce near-optimal objective function values. Lack of organization in models like these can be closely related to practical non-identifiability, in which parameter estimates cannot be uniquely determined given the data. In our profile likelihood analysis of all 28 fitted parameters, we found that approximately half were practically identifiable, while the remainder showed broad, flat profiles, indicating limited sensitivity to perturbation. The full results are shown in Supplementary Figure S4, which plots each parameter against its objective function profile, maximum likelihood estimate, and deviation from the original baseline from Kocieniewski and Lipniacki (2013). Although this lack of uniqueness highlights the need for additional experimental data to fully constrain the system, it is important to note that model predictions remained consistent across different parameter sets. Thus, despite non-identifiable parameters, the model exhibits robustness, supporting the reliability of the conclusions drawn from our PyBioNetFit-enabled analysis.

Our results illustrate several advantages of a PyBioNetFit-enabled workflow, including potential for straightforward reusability of qualitative observations, once formalized as BPSL statements, and reproducibility of complex workflows. PyBioNetFit job setup files consist of easily shared plain-text files capturing data, models, and algorithms in standardized formats. Our results also reinforce the value of qualitative data. These data can be used in model parameterization with UQ, in multiple ways. Looking ahead, this study serves as a template for enhancing model parameterization pipelines in contexts where qualitative data are available to complement (limited) quantitative data.

## 4.1 Limitations and Future Directions

While our study demonstrates the utility of PyBioNetFit for automated parameterization with both qualitative and quantitative data, several limitations should be acknowledged. First, scalability remains a practical challenge. Although PyBioNetFit performs well for models of moderate size, applying it to models with hundreds or thousands of parameters, especially those that are high-dimensional, will become too computationally expensive. Future work may benefit from hybrid optimization strategies that combine global search with local refinement to improve efficiency.

Secondly, while the Biological Property Specification Language (BPSL) is well suited to encoding rank orderings, inequalities at fixed times, and cross-condition comparisons, it does not currently support constraints on more complex temporal features, such as peak times. Extending BPSL to represent such properties, as well as improving its ability to incorporate metadata (e.g., experimental conditions, cell types), would expand its applicability.

Third, BPSL has potential beyond parameter estimation. For example, BPSL statements could be leveraged in model-checking workflows to systematically document a model's consistency with qualitative biological knowledge. In addition, advances in natural language processing, particularly large language models, could enable automated extraction of BPSL statements from full-text articles, further broadening their utility in parameterization and uncertainty quantification workflows.

Taken together, these considerations highlight important directions for future research. Our demonstration emphasizes that while current approaches are powerful, there is significant scope for enhancing both scalability and expressivity, thereby enabling broader adoption of hybrid data workflows in systems biology.

## 5 Conflict of Interest

The authors declare that the research was conducted in the absence of any commercial or financial relationships that could be construed as a potential conflict of interest.

## 6 Funding

We acknowledge support from the National Institute of General Medical Sciences of the National Institutes of Health through grant R01GM111510, and support from the National Institute of Allergy and Infectious Diseases of the National Institutes of Health through grant R01AI153617. We also received support from the Center for Nonlinear Studies and the Laboratory-Directed Research and Development Program at Los Alamos National Laboratory. This study benefitted from Northern Arizona University's Monsoon computer cluster, which is supported by Arizona's Technology and Research Initiative Fund, and Los Alamos National Laboratory's Darwin computer cluster, which is supported by the Computational Systems and Software Environment subprogram of the Advanced Simulation and Computing program at Los Alamos National Laboratory.

## 7 Acknowledgments

We thank Dr. T. Lipniacki for helpful discussions.

# 9 Tables

| Parameter | Original Parameter Value | Parameter Units | Parameter Description |
| --- | --- | --- | --- |
| $c_1L$ | $2.0\text{x}10^{-2}$ | Not Provided | c1 value after stimulation with ligand |

| $c_2$ | $2.0x10^{-7}$ | $(mcls\ x\ s)^{-1}$ | EGFR receptor dimerization dueligand |
|---|---|---|---|
| $t_1$ | $1.0x10^{2}$ | $s^{-1}$ | EGFR subunits transphosphorylation in EGFR dimer |
| $d_3$ | $1.0x10^{-3}$ | $s^{-1}$ | degradation of ligand bound dimer complexes |
| $b_1$ | $4.0x10^{-8}$ | $(mcls\ x\ s)^{-1}$ | association of phosphorylated receptor EGFR subunits with SOS1 |
| $n_1$ | $2.0x10^{-3}$ | $s^{-1}$ | disassociation of EGFR receptor subunits from SOS1 |
| $b_2$ | $1.0x10^{-5}$ | $(mcls\ x\ s)^{-1}$ | MEK1 homodimer formation |
| $n_2$ | $1.0x10^{-3}$ | $s^{-1}$ | MEK1 dimer dissociation |
| $b_3$ | $1.0x10^{-5}$ | $(mcls\ x\ s)^{-1}$ | MEK2 homodimer formation |
| $n_3$ | $3.0x10^{-2}$ | $s^{-1}$ | MEK2 homodimer dissociation |
| $b_4$ | $1.0x10^{-5}$ | $(mcls\ x\ s)^{-1}$ | MEK1 and MEK2 heterodimer formation |
| $n_4$ | $1.0x10^{-3}$ | $s^{-1}$ | MEK1 and MEK2 heterodimer dissociation |
| $a_1$ | $1.5x10^{-7}$ | $(mcls\ x\ s)^{-1}$ | activation of Ras by EGFR SOS1 complex (exchange of GDP for GTP) |
| $i_1$ | $2.0x10^{-2}$ | $s^{-1}$ | inactivation of RAS (hydrolysis of bound GTP to GDP) |
| $a_2$ | $4.0x10^{-8}$ | $(mcls\ x\ s)^{-1}$ | activation of RAF by RAS − GTP |
| $i_2$ | $1.0x10^{-2}$ | $s^{-1}$ | inactivation of RAF |
| $p_1$ | $1.5x10^{-7}$ | $(mcls\ x\ s)^{-1}$ | phosphorylation of MEK1 and MEK2 on the activation sites by RAF |
| $u_1$ | $5.0x10^{-3}$ | $s^{-1}$ | dephosphorylation of MEK1 and MEK2 on the activation sites |
| $p_2a$ | $1.0x10^{-6}$ | $(mcls\ x\ s)^{-1}$ | phosphorylation of ERK on the activation sites by MEK1 |
| $p_2b$ | $5.0x10^{-6}$ | $(mcls\ x\ s)^{-1}$ | phosphorylation of ERK on the activation sites by MEK2 |

| $u_2$ | $2.0x10^{-2}$ | $s^{-1}$ | dephosphorylation of ERK on the activation sites |
|---|---|---|---|
| $p_3$ | $2.0x10^{-9}$ | $(mcls\ x\ s)^{-1}$ | feedback phosphorylation of SOS1 by active ERK |
| $u_3$ | $1.0x10^{-3}$ | $(mcls\ x\ s)^{-1}$ | dephosphorylation of the SOS1 feedback site |
| $p_4$ | $1.2x10^{-9}$ | $(mcls\ x\ s)^{-1}$ | feedback phosphorylation of MEK1 by ERK |
| $u_4$ | $2.0x10^{-4}$ | $(mcls\ x\ s)^{-1}$ | dephosphorylation of the MEK1 feedback site (Thr292) |
| $b_5$ | $4.0x10^{-9}$ | $(mcls\ x\ s)^{-1}$ | PHP phosphatse binding to Thr292p of MEK1 |
| $n_5$ | $2.0x10^{-4}$ | $s^{-1}$ | dissociation of PHP phosphatase from Thr292p of MEK1 |
| $u_5$ | $2.0x10^{1}$ | $s^{-1}$ | dephosphorylation of MEK1 and MEK2 on the activation sites |
| $s_1$ * | 2.5 | $s^{-1}$ | EGFR receptor subunit constitutive production |
| $d_1$ * | $5.0x10^{-6}$ | $s^{-1}$ | EGFR receptor subunit constitutive degredation |
| $s_2$ * | 1.0 | $s^{-1}$ | Sos1 constitutive production |
| $d_2$ * | $5.0x10^{-6}$ | $s^{-1}$ | Sos1 constitutive degredation |
| $c_1$ * | $2.0x10^{-2}$ | $N/A$ | formation of EGFR receptor subunit − ligand complex **Starting Signal** |
| $c1_{init}$ * | 0.0 | $(mcls\ x\ s)^{-1}$ | $c_1$ value after simulation with ligand |
| MEK1_fraction* | $6.7x10^{-1}$ | $N/A$ | MEK1 fraction of total MEKs content |
| X* | 5.0 | $N/A$ | MEK2 MEK1 kinase activity ratio |
| $EGFR_0$* | $5.0x10^{5}$ | *Copy Number* | initial EGFR level |

| $SOS1_0$* | $2.0x10^5$ | *Copy Number* | initial Sos1 level |
|---|---|---|---|
| $RAS_0$* | $5.0x10^5$ | *Copy Number* | initial RAS level |
| $RAF_0$* | $5.0x10^5$ | *Copy Number* | initial RAF level |
| $MEK_tot_0$* | $2.0x10^5$ | *Copy Number* | initial MEK1 and MEK2 total level |
| $MEK1_0$* | $1.34x10^5$ | *Copy Number* | initial MEK1 level |
| $MEK1_T292p_0$* | 0.0 | *Copy Number* | initial MEK1 − Thr292p level |
| $MEK2_0$* | $6.6x10^4$ | *Copy Number* | initial MEK2 level |
| $ERK_0$* | $3.0x10^6$ | *Copy Number* | initial combined ERK1 and ERK2 level |
| $PHP_MEK_0$* | $3.0x10^6$ | *Copy Number* | initial PHP level |

**Table 1:** Original parameter value estimates of the MEK isoform model determined by Kocieniewski and Lipniacki (2013). All original 46 parameters are shown, including a brief description of their function within the MEK isoform model. Parameters marked with a single asterisk (*) denote parameters that were fixed during PyBioNetFit's automated parameterization of the model.

| Parameter | Best-fit Value | Optimization Bounds **[low, high]** | Bayesian UQ Bounds **[low, high]** |
|---|---|---|---|
| $c_1L$ | $7.4x10^{-3}$ | $[9.0x10^{-4}, 4.0x10^{-2}]$ | N/A |
| $c_2$ | $9.3x10^{-9}$ | $[9.0x10^{-9}, 4.0x10^{-6}]$ | N/A |
| $t_1$ | $9.9x10^1$ | $[99, 101]$ | N/A |
| $d_3$ | $2.0x10^{-3}$ | $[8.0x10^{-5}, 3.0x10^{-2}]$ | $[1.0x10^{-5}, 1.0x10^{-1}]$ |
| $b_1$ | $2.4x10^{-8}$ | $[2.0x10^{-9}, 6.0x10^{-7}]$ | N/A |
| $n_1$ | $6.5\text{x}10^{-4}$ | $[9.0\text{x}10^{-5}, 4.0\text{x}10^{-2}]$ | N/A |
| $b_2$ | $4.2\text{x}10^{-6}$ | $[8.0\text{x}10^{-7}, 3.0\text{x}10^{-4}]$ | N/A |
| $n_2$ | $2.6\text{x}10^{-4}$ | $[8.0\text{x}10^{-5}, 3.0\text{x}10^{-2}]$ | N/A |
| $b_3$ | $9.9\text{x}10^{-5}$ | $[8.0\text{x}10^{-7}, 3.0\text{x}10^{-4}]$ | N/A |
| $n_3$ | $8.0\text{x}10^{-2}$ | $[1.0\text{x}10^{-3}, 5.0\text{x}10^{-1}]$ | N/A |
| $b_4$ | $2.9\text{x}10^{-4}$ | $[8.0\text{x}10^{-7}, 3.0\text{x}10^{-4}]$ | N/A |
| $n_4$ | $3.5\text{x}10^{-3}$ | $[8.0\text{x}10^{-5}, 3.0\text{x}10^{-2}]$ | N/A |
| $a_1$ | $2.3\text{x}10^{-7}$ | $[1.27\text{x}10^{-8}, 1.29\text{x}10^{-6}]$ | N/A |
| $i_1$ | $3.9\text{x}10^{-1}$ | $[9.0\text{x}10^{-4}, 4.0\text{x}10^{-1}]$ | N/A |
| $a_2$ | $3.3\text{x}10^{-8}$ | $[2.0\text{x}10^{-9}, 6.0\text{x}10^{-7}]$ | N/A |
| $i_2$ | $2.7\text{x}10^{-2}$ | $[8.0\text{x}10^{-4}, 3.0\text{x}10^{-1}]$ | N/A |
| $p_1$ | $3.5\text{x}10^{-6}$ | $[8.0\text{x}10^{-9}, 4.0\text{x}10^{-6}]$ | N/A |
| $u_1$ | $9.5\text{x}10^{-4}$ | $[3.0\text{x}10^{-4}, 7.0\text{x}10^{-2}]$ | N/A |
| $p_2a$ | $5.2\text{x}10^{-7}$ | $[8.0\text{x}10^{-8}, 3.0\text{x}10^{-5}]$ | N/A |
| $p_2b$ | $2.3\text{x}10^{-5}$ | $[1.0\text{x}10^{-6}, 5.0\text{x}10^{-4}]$ | N/A |
| $u_2$ | $6.7\text{x}10^{-2}$ | $[9.0\text{x}10^{-4}, 4.0\text{x}10^{-1}]$ | N/A |
| $p_3$ | $1.1\text{x}10^{-9}$ | $[9.0\text{x}10^{-11}, 4.0\text{x}10^{-8}]$ | N/A |
| $u_3$ | $2.6\text{x}10^{-4}$ | $[8.0\text{x}10^{-5}, 3.0\text{x}10^{-2}]$ | $[1.0\text{x}10^{-5}, 1.0\text{x}10^{-1}]$ |
| $p_4$ | $6.5\text{x}10^{-10}$ | $[9.0\text{x}10^{-11}, 3.0\text{x}10^{-8}]$ | N/A |

| $u_4$ | $3.2x10^{-4}$ | $[9.0x10^{-6}, 4.0x10^{-3}]$ | N/A |
|---|---|---|---|
| $b_5$ | $1.7x10^{-8}$ | $[2.0x10^{-10}, 6.0x10^{-8}]$ | N/A |
| $n_5$ | $3.9x10^{-3}$ | $[9.0x10^{-6}, 4.0x10^{-3}]$ | N/A |
| $u_5$ | $1.9x10^{1}$ | $[19, 21]$ | N/A |
| $Scale_{pEGFR}$* | $1.1x10^{-4}$ | $[3.0x10^{-5}, 6.0x10^{-5}]$ | $[3.0x10^{-5}, 6.0x10^{-5}]$ |
| $Scale_{pERK}$* | $3.2x10^{-6}$ | $[2.0x10^{-6}, 5.0x10^{-6}]$ | $[2.0x10^{-6}, 5.0x10^{-6}]$ |
| $Scale_{pSOS1}$* | $1.1x10^{-4}$ | $[9.0x10^{-5}, 3.0^{-4}]$ | $[9.0x10^{-5}, 3.0^{-4}]$ |
| $sigma$ ** | 1.1 | N/A | $[1.0x10^{-1}, 1.0x10^{1}]$ |

**Table 2:** Best-fit parameter estimates**.** This table lists the 31 parameters selected for model calibration using PyBioNetFit. Each row indicates the name of a parameter, the best-fit value obtained from a global fit, and parameter bounds used in optimization and Bayesian UQ (if applicable). Parameters marked with a single asterisk (*) are scaling factors added specifically for the WT model to allow model outputs to be compared to published measurements reported in arbitrary units (AU). Parameters with a double asterisk (**) are special hyper sampling parameters used only in Bayesian UQ simulations needed for PyBioNetFit's Adaptive MCMC algorithm and not originally included in the model by Kocieniewski and Lipniacki (2013) .

| | $ESS_{Bulk}$ | $ESS_{Tail}$ | $\hat{R}-1$ |
|---|---|---|---|
| $d_3$ | 6160 | 8620 | $1.8x10^{-4}$ |
| $u_3$ | 2150 | 2120 | $6.0x10^{-4}$ |

**Table 3.** Sampling convergence metrics. For the rate constants $d_3$ and $u_3$, the table reports the *bulk* and *tail* Effective Sample Size (ESS), as well as the quantity $\hat{R}-1$, in which $\hat{R}$ is a convergence diagnostic derived from the potential scale reduction factor and computed using the rstan R package. $ESS_{Bulk}$ evaluates the effective number of independent samples for the central bulk of the posterior distribution, whereas $ESS_{Tail}$ reflects the stability of sampling in the distribution tails. High ESS values across both metrics indicate efficient sampling with low autocorrelation. The $\hat{R}-1$ values are close to zero, suggesting convergence of the chains and well-mixed posterior estimates.

## 10 Figure Legends

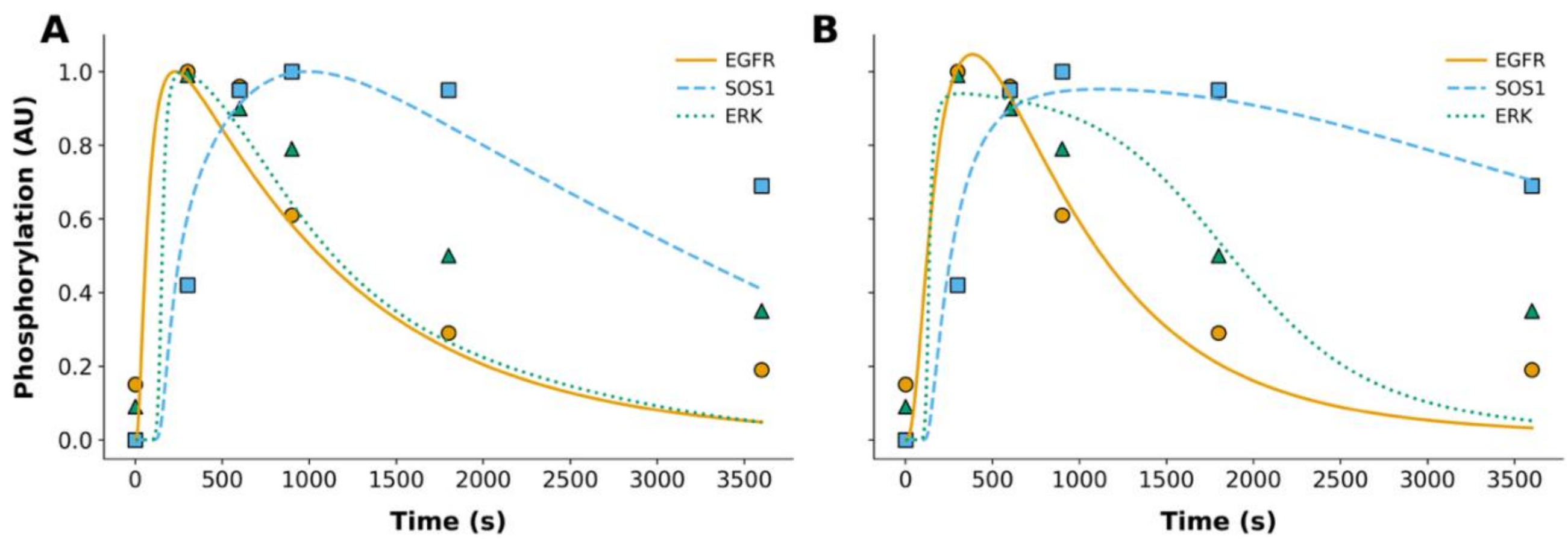

**Figure 1:** Comparison of model-generated best-fit trajectories with experimental phosphorylation data in WT conditions. Each panel displays simulated trajectories for phosphorylated EGFR (solid orange), SOS1 (dashed light blue), and ERK (dotted green), plotted alongside their respective experimental data points (colored to match model predictions). All trajectories are based on MLEs for parameters. Panel A shows the model outputs derived from the original parameterization of Kocieniewski and Lipniacki (2013). Panel B presents the MLEs obtained using PyBioNetFit, as described in Materials and Methods. Phosphorylation values are reported in arbitrary units (AU). Experimental data points are consistent between panels.

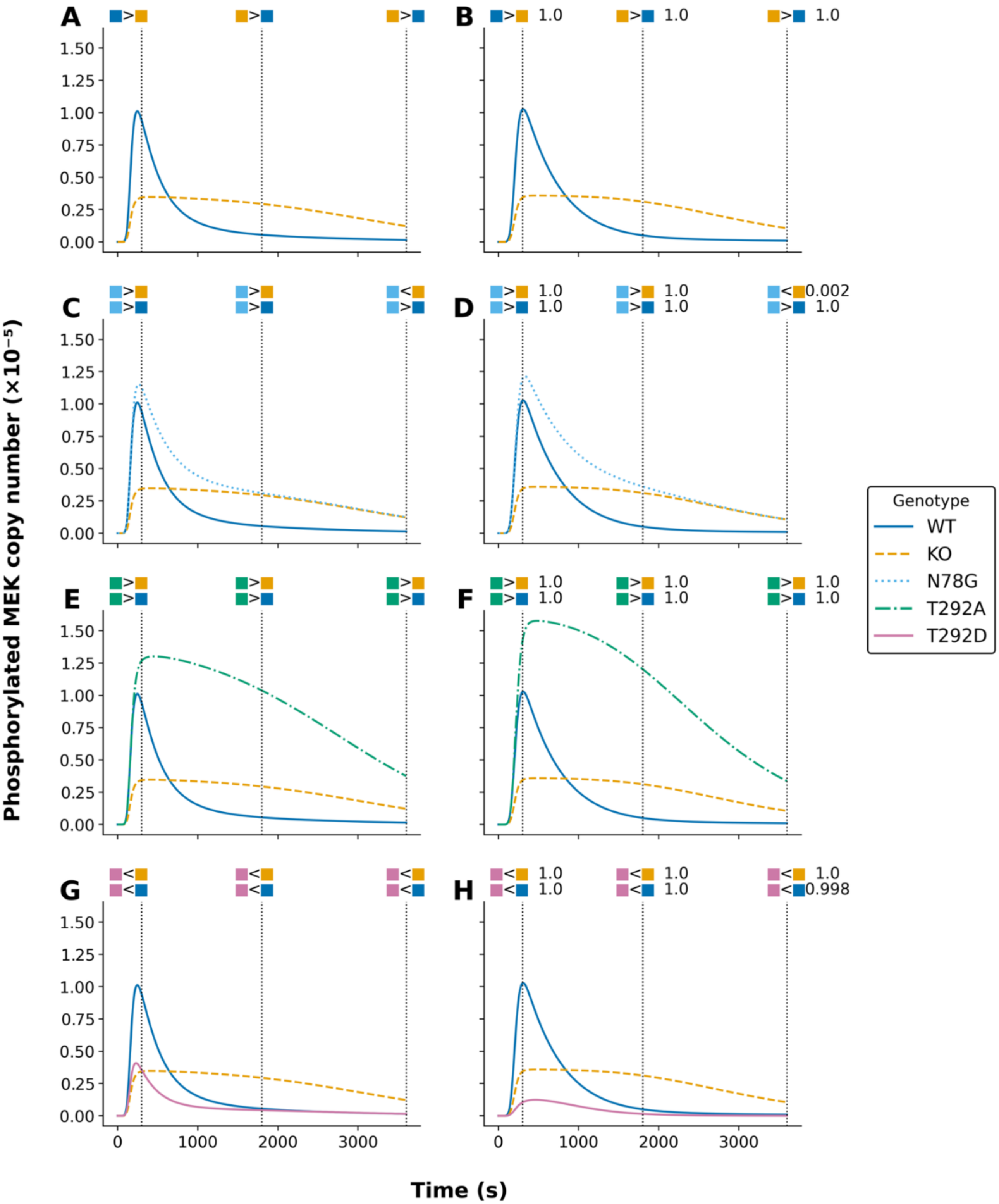

**Figure 2:** Comparison of model-predicted phosphorylated MEK trajectories under different parameterizations for five model variants. Each panel illustrates trajectories for phosphorylated MEK (in molecules $\times 10^{-5}$) for indicated models. Curves are color-coded by model variant: wild type (WT, solid blue), knockout (KO, dashed orange), N78G (dotted light blue), T292A (dashed-dotted green), and T292D (solid pink). The left panels (A, C, E, G) display model outputs derived from the original parameterization of Kocieniewski and Lipniacki (2013). The right panels (B, D, F, H) show outputs

based on the parameterization obtained using PyBioNetFit, as described in Materials and Methods. Constraints used in parameterization are represented as colored glyphs above each panel and match the corresponding model variant by color. Vertical black dotted lines indicate the time points at which these constraints apply: 300, 1800, and 3600 seconds. For panels B, D, F, and H (PyBioNetFit parameterization), numerical annotations indicate the fraction of accepted MCMC samples satisfying each constraint.

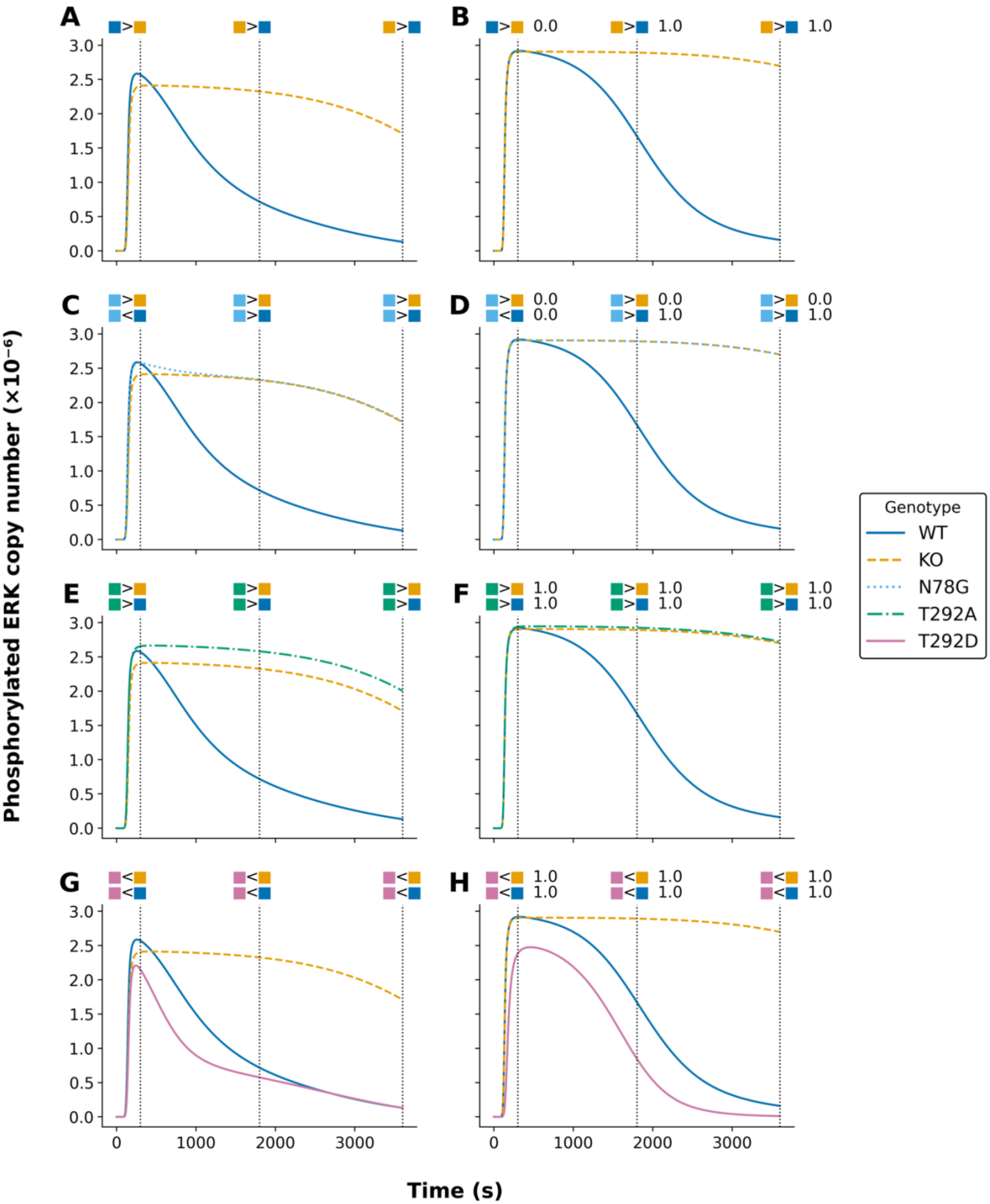

**Figure 3:** Comparison of model-predicted phosphorylated ERK trajectories under different parameterizations for five model variants. Each panel illustrates trajectories for phosphorylated ERK (in molecules ×10$^{-6}$) for indicated models. Curves are color-coded by model variant: wild type (WT, solid blue), knockout (KO, dashed orange), N78G (dotted light blue), T292A (dashed-dotted green), and T292D (solid pink). The left panels (A, C, E, G) display model outputs derived from the original

parameterization of Kocieniewski and Lipniacki (2013). The right panels (B, D, F, H) show outputs based on the parameterization obtained using PyBioNetFit, as described in Materials and Methods. Constraints used in parameterization are represented as colored glyphs above each panel and match the corresponding model variant by color. Vertical black dotted lines indicate the time points at which these constraints apply: 300, 1800, and 3600 seconds. For panels B, D, F, and H (PyBioNetFit parameterization), numerical annotations indicate the fraction of accepted MCMC samples satisfying each constraint.

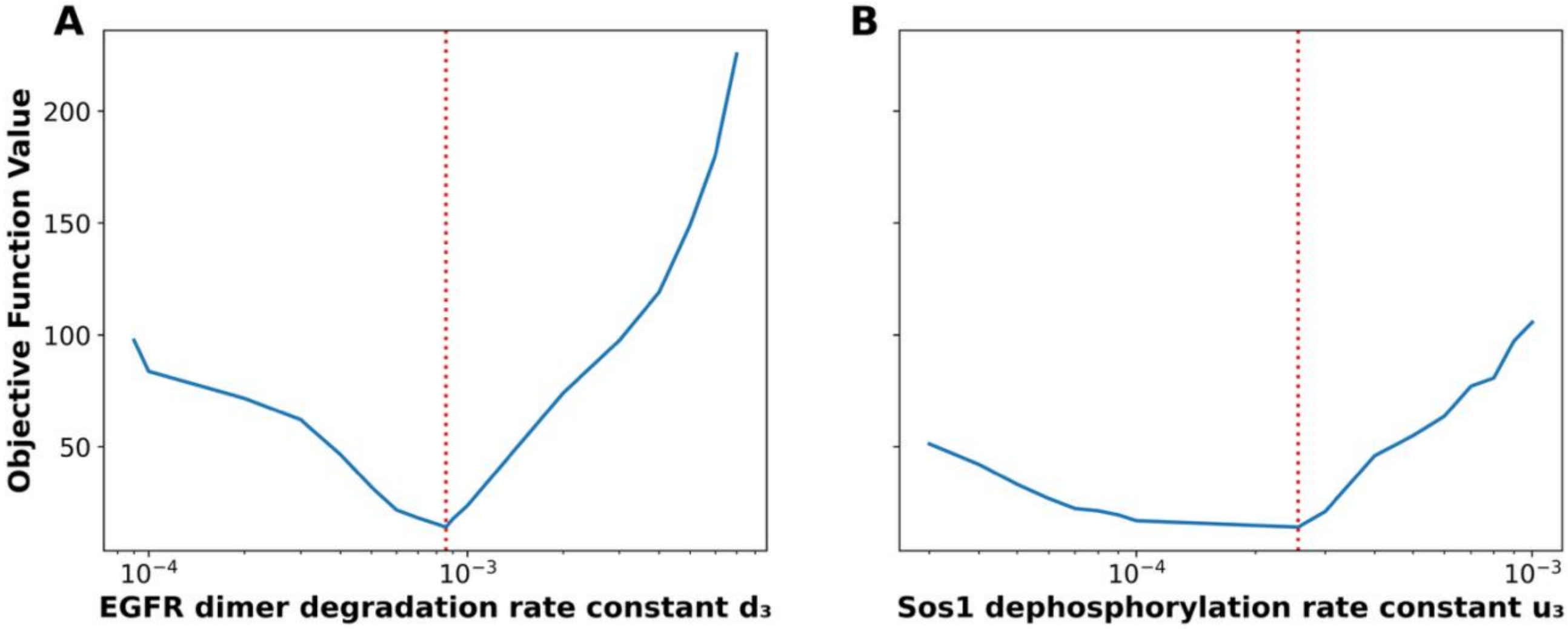

**Figure 4:** Profile likelihood analysis. Panel A displays the profile likelihood for the EGFR dimer degradation rate constant $d_3$, and Panel B displays the profile likelihood for the SOS1 dephosphorylation rate constant $u_3$. Each solid blue curve reports the minimum objective function value found in PyBioNetFit-enabled optimization when the indicated parameter, $d_3$ or $u_3$, is held fixed at the value indicated on the horizontal axis and the remaining parameters are allowed to vary. The vertical red dashed line in each panel indicates the MLE.

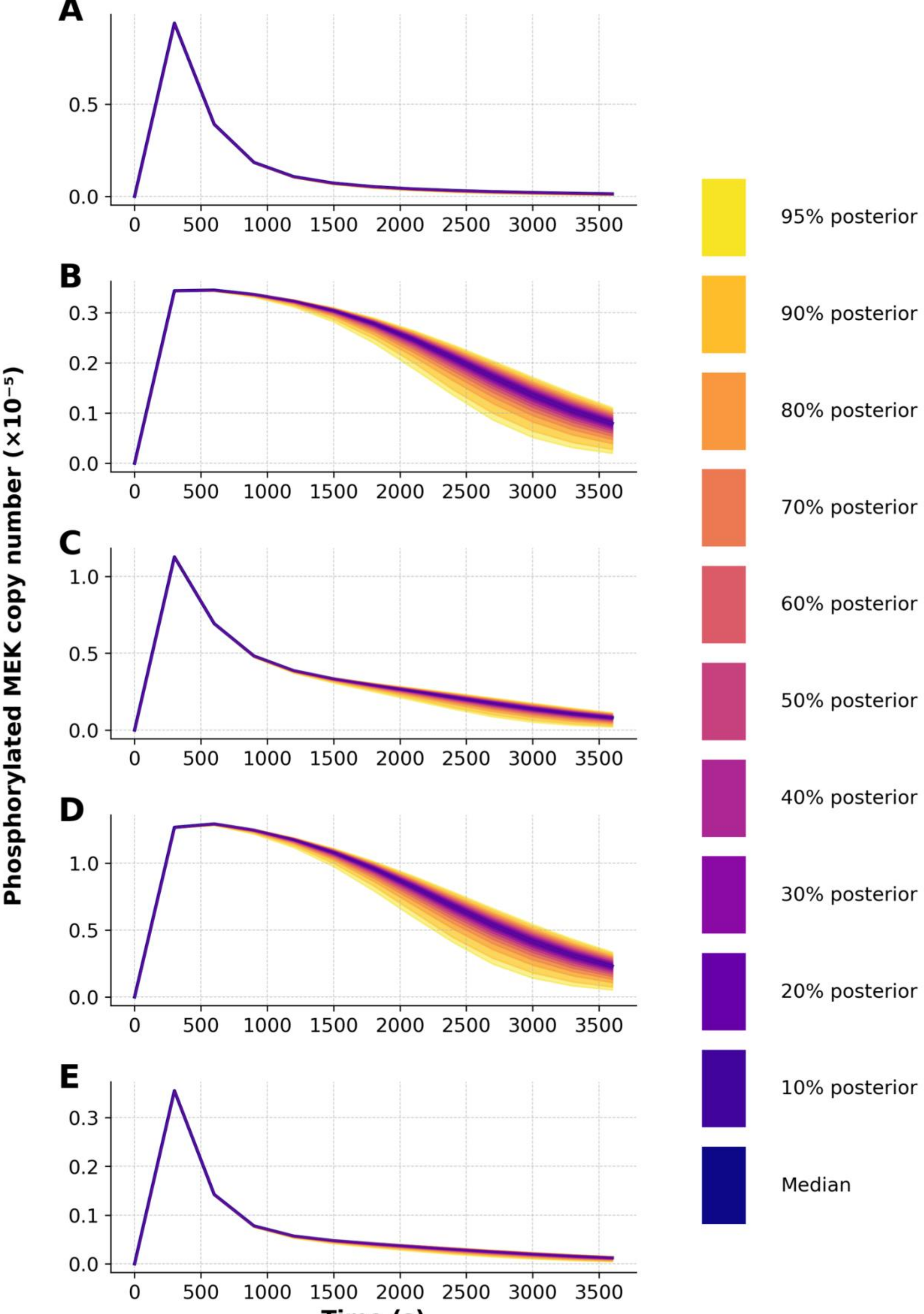

**Figure 5:** Posterior predictive distributions of phosphorylated MEK copy number for each of the five model variants. Each panel corresponds to one model, as follows: (A) WT, (B) KO, (C) N78G, (D) T292A, and (E) T292D. The dark blue lines represent medians. Shaded bands indicate credible intervals from 10% to 95%, with the outermost band corresponding to the 95% credible interval. The distribution shown at each time was obtained by propagating parametric uncertainty through simulations, without injection of noise from our noise model.

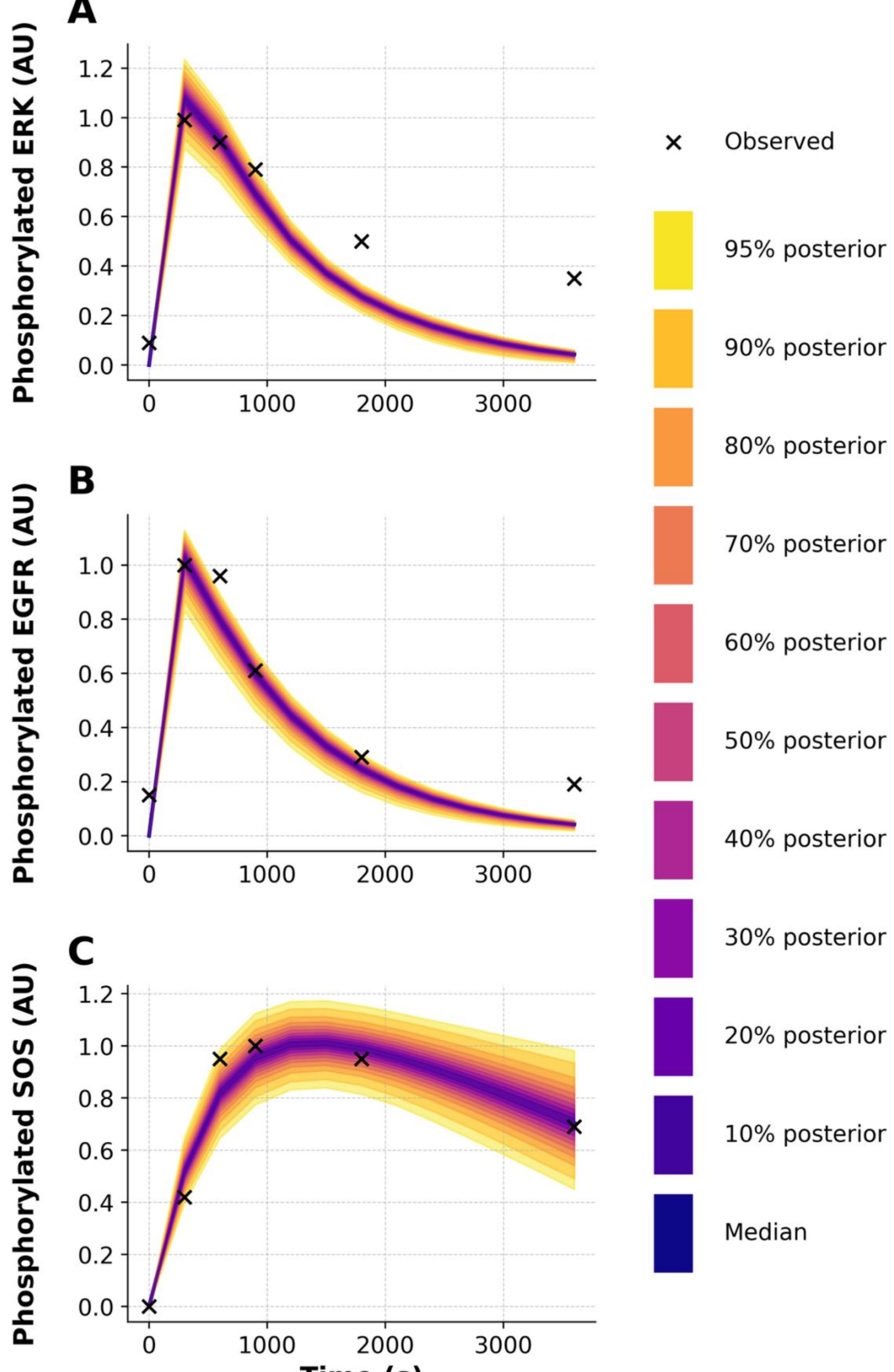

**Figure 6:**

Posterior predictive distributions of phosphorylation outputs of the WT model: (A) phosphorylated ERK, (B) phosphorylated EGFR, and (C) phosphorylated SOS1, all expressed in arbitrary units (AU). The distribution shown at each time point was obtained by propagating parametric uncertainty through simulations, without injection of noise from our noise model. Experimental data are included in each panel. Dark blue lines represent median trajectories, and colored bands indicate credible intervals from 10% to 95%, with the outermost band corresponding to the 95% credible interval.

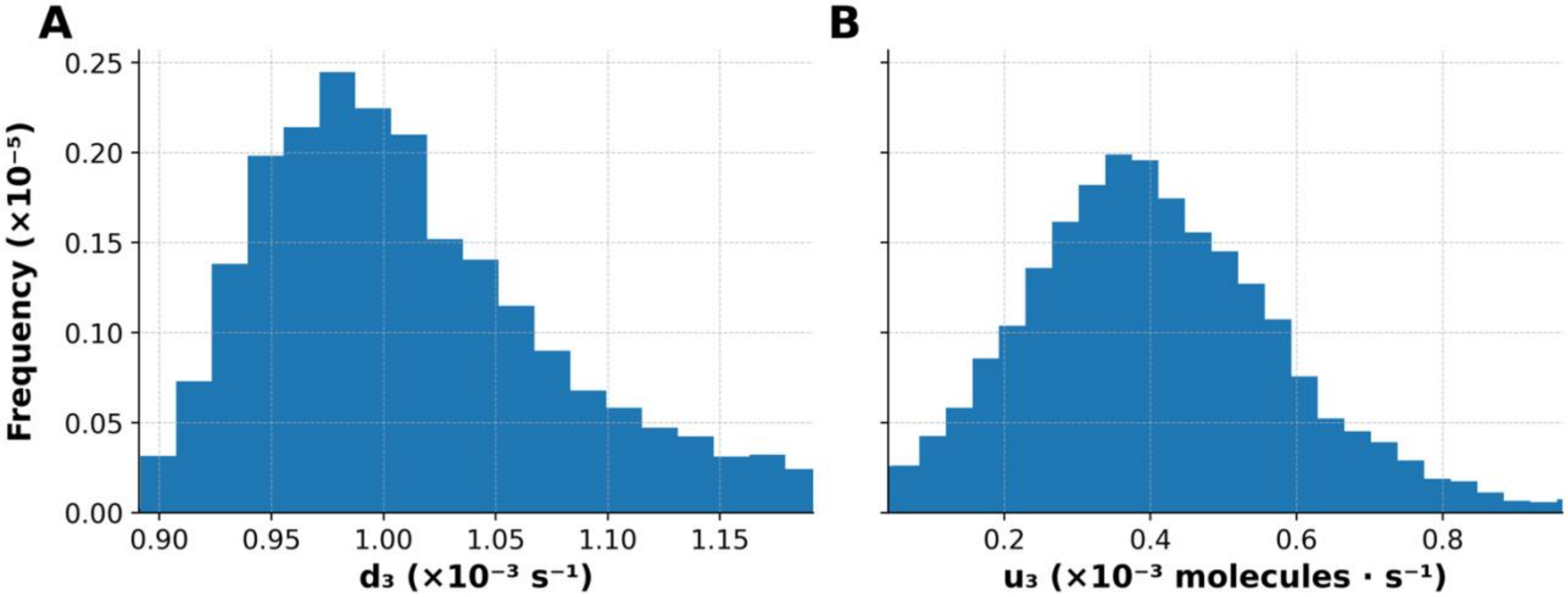

**Figure 7:** Marginal posterior distributions for the rate constants (A) $d_3$ and (B) $u_3$. These histograms were derived from samples generated using the adaptive MCMC sampler implemented in PyBioNetFit, as described in Materials and Methods. Parametric uncertainty is characterized by histogram width.

## 11 Supplementary Material

Supplementary Material includes Supplementary Tables 1–6 and Supplementary Figures 1–3.

## 12 Data Availability Statement

The code, datasets, and job setup files used in this study can be found in the PyBioNetFit GitHub repository (https://github.com/lanl/PyBNF).

# Supplementary Material

## 1 Glossary of Terms and Abbreviations

**BPSL (Biological Property Specification Language):** A formal language used to encode qualitative biological constraints (e.g., comparisons of time-course values between species or conditions).

**Constraint (C#):** A numbered qualitative rule (C1, C2, etc.) expressed in BPSL, used to test whether model simulations match expected biological behaviors. Constraint glossaries for each model are provided in Supplemental Tables 1–5.

**DE (Differential Evolution):** A global optimization algorithm used in PyBioNetFit for parameter estimation.

**Isoform Models (WT, KO, N78G, T292A, T292D):** Variants of the MEK signaling pathway model.

- **WT:** Wild-type model.
- **KO:** Knockout model (gene disrupted).
- **N78G:** MEK isoform model with the N78G mutation.
- **T292A:** MEK isoform model with the T292A mutation.
- **T292D:** MEK isoform model with the T292D mutation.

**MLE (Maximum Likelihood Estimation):** Statistical method for estimating parameters by maximizing the likelihood of the observed data given the model.

**MCMC (Markov Chain Monte Carlo):** Sampling method used to explore posterior distributions in Bayesian inference.

**Objective Function (SOS):** In PyBioNetFit, the Sum of Squares (SOS) objective function measures the difference between model predictions and data; lower scores indicate better fits.

**pERK1/2:** Phosphorylated ERK1/2, a key downstream signaling protein measured in the models.

**pMEK (MEK_pRDS):** Phosphorylated MEK (measured as relative densitometry signal, RDS) in experimental datasets and model outputs.

**pSOS1, pEGFR, pERK:** Phosphorylated forms of signaling proteins SOS1, EGFR, and ERK, respectively; used for quantitative comparison between experiment and simulation.

**Profile Likelihood Analysis:** A method to assess parameter identifiability by varying one parameter at a time while re-optimizing others, producing likelihood profiles.

**PyBioNetFit:** Software framework used for parameterization and uncertainty quantification of biological models using quantitative data and qualitative constraints.

**RMSE (Root Mean Squared Error):** A measure of model fit, quantifying the average difference between model outputs and experimental data; lower values indicate better fits.

## 2 Supplementary Data

The source code used in this study can be found in the PyBioNetFit GitHub repository (https://github.com/lanl/PyBNF). Within this repository, the datasets and job setup files used in this study for maximum likelihood estimation and Bayesian inference can be found here: https://github.com/lanl/PyBNF/tree/master/examples/Miller2025_MEK_Isoforms

## 3 Supplementary Tables and Figures

| Constraint # | WT Model |
|---|---|
| C1 | WT.MEK_pRDS at time=300 > WT.MEK_pRDS at time=1800 |
| C2 | WT.MEK_pRDS at time=1800 > WT.MEK_pRDS at time=3600 |
| C3 | WT.MEK_pRDS at time=300 > WT.MEK_pRDS at time=3600 |
| C4 | WT.MEK_pRDS at time=300 > KO.MEK_pRDS at time=300 |
| C5 | WT.MEK_pRDS at time=1800 < KO.MEK_pRDS at time=1800 |
| C6 | WT.MEK_pRDS at time=3600 < KO.MEK_pRDS at time=3600 |
| C7 | WT.MEK_pRDS at time=300 < N78G.MEK_pRDS at time=300 |
| C8 | WT.MEK_pRDS at time=1800 < N78G.MEK_pRDS at time=1800 |
| C9 | WT.MEK_pRDS at time=3600 < N78G.MEK_pRDS at time=3600 |
| C10 | WT.MEK_pRDS at time=300 < T292A.MEK_pRDS at time=300 |

| C11 | WT.MEK_pRDS at time=1800 < T292A.MEK_pRDS at time=1800 |
|---|---|
| C12 | WT.MEK_pRDS at time=3600 < T292A.MEK_pRDS at time=3600 |
| C13 | WT.MEK_pRDS at time=300 > T292D.MEK_pRDS at time=300 |
| C14 | WT.MEK_pRDS at time=1800 > T292D.MEK_pRDS at time=1800 |
| C15 | WT.MEK_pRDS at time=3600 > T292D.MEK_pRDS at time=3600 |
| C16 | WT.pERK1_2 at time=300 > WT.pERK1_2 at time=1800 |
| C17 | WT.pERK1_2 at time=1800 > WT.pERK1_2 at time=3600 |
| C18 | WT.pERK1_2_ at time=300 > WT.pERK1_2 at time=3600 |
| C19 | WT.pERK1_2 at time=300 > KO.pERK1_2 at time=300 |
| C20 | WT.pERK1_2 at time=1800 < KO.pERK1_2 at time=1800 |
| C21 | WT.pERK1_2 at time=3600 < KO.pERK1_2 at time=3600 |
| C22 | WT.pERK1_2 at time=300 > N78G.pERK1_2 at time=300 |
| C23 | WT.pERK1_2 at time=1800 < N78G.pERK1_2 at time=1800 |
| C24 | WT.pERK1_2 at time=3600 < N78G.pERK1_2 at time=3600 |
| C25 | WT.pERK1_2 at time=300 < T292A.pERK1_2 at time=300 |
| C26 | WT.pERK1_2 at time=1800 < T292A.pERK1_2 at time=1800 |
| C27 | WT.pERK1_2 at time=3600 < T292A.pERK1_2 at time=3600 |
| C28 | WT.pERK1_2 at time=300 > T292D.pERK1_2 at time=300 |
| C29 | WT.pERK1_2 at time=1800 > T292D.pERK1_2 at time=1800 |
| C30 | WT.pERK1_2 at time=3600 > T292D.pERK1_2 at time=3600 |

**Supplemental Table 1 (WT constraint glossary):**
This table provides a glossary for the constraint labels (C1–C30) used in Supplemental Table 6 for the WT model. Each entry is a BPSL statement that formalizes a qualitative observation. Time points in the statements are given in seconds.

| Constraint # | KO Model |
|---|---|
| C1 | KO.MEK_pRDS at time=300 > KO.MEK_pRDS at time=1800 |
| C2 | KO.MEK_pRDS at time=300 > KO.MEK_pRDS at time=3600 |
| C3 | KO.MEK_pRDS at time=1800 > KO.MEK_pRDS at time=3600 |
| C4 | KO.MEK_pRDS at time=300 < N78G.MEK_pRDS at time=300 |
| C5 | KO.MEK_pRDS at time=1800 < N78G.MEK_pRDS at time=1800 |
| C6 | KO.MEK_pRDS at time=3600 > N78G.MEK_pRDS at time=3600 |
| C7 | KO.MEK_pRDS at time=300 < T292A.MEK_pRDS at time=300 |
| C8 | KO.MEK_pRDS at time=1800 < T292A.MEK_pRDS at time=1800 |
| C9 | KO.MEK_pRDS at time=3600 < T292A.MEK_pRDS at time=3600 |
| C10 | KO.MEK_pRDS at time=300 > T292D.MEK_pRDS at time=300 |
| C11 | KO.MEK_pRDS at time=1800 > T292D.MEK_pRDS at time=1800 |

| C12 | KO.MEK_pRDS at time=3600 > T292D.MEK_pRDS at time=3600 |
|---|---|
| C13 | KO.pERK1_2 at time=300 > KO.pERK1_2at time=1800 |
| C14 | KO.pERK1_2 at time=1800 > KO.pERK1_2 at time=3600 |
| C15 | KO.pERK1_2 at time=300 > KO.pERK1_2 at time=3600 |
| C16 | KO.pERK1_2 at time=300 < N78G.pERK1_2 at time=300 |
| C17 | KO.pERK1_2 at time=1800 < N78G.pERK1_2 at time=1800 |
| C18 | KO.pERK1_2 at time=3600 < N78G.pERK1_2 at time=3600 |
| C19 | KO.pERK1_2 at time=300 < T292A.pERK1_2 at time=300 |
| C20 | KO.pERK1_2 at time=1800 < T292A.pERK1_2 at time=1800 |
| C21 | KO.pERK1_2 at time=3600 < T292A.pERK1_2 at time=3600 |
| C22 | KO.pERK1_2 at time=300 > T292D.pERK1_2 at time=300 |
| C23 | KO.pERK1_2 at time=1800 > T292D.pERK1_2 at time=1800 |
| C24 | KO.pERK1_2 at time=3600 > T292D.pERK1_2 at time=3600 |

**Supplemental Table 2 (KO constraint glossary):**
This table provides a glossary for the constraint labels (C1–C24) used in Supplemental Table 6 for the KO model. Each entry is a BPSL statement that formalizes a qualitative observation. Time points in the statements are given in seconds.

\

| Constraint # | N78G Model |
|---|---|
| C1 | N78G.MEK_pRDS at time=300 > N78G.MEK_pRDS at time=1800 |
| C2 | N78G.MEK_pRDS at time=1800 > N78G.MEK_pRDS at time=3600 |
| C3 | N78G.MEK_pRDS at time=300 > N78G.MEK_pRDS at time=3600 |
| C4 | N78G.MEK_pRDS at time=300 < T292A.MEK_pRDS at time=300 |
| C5 | N78G.MEK_pRDS at time=1800 < T292A.MEK_pRDS at time=1800 |
| C6 | N78G.MEK_pRDS at time=3600 < T292A.MEK_pRDS at time=3600 |
| C7 | N78G.MEK_pRDS at time=300 > T292D.MEK_pRDS at time=300 |
| C8 | N78G.MEK_pRDS at time=1800 > T292D.MEK_pRDS at time=1800 |
| C9 | N78G.MEK_pRDS at time=3600 > T292D.MEK_pRDS at time=3600 |
| C10 | N78G.pERK1_2 at time=300 > N78G.pERK1_2 at time=1800 |
| C11 | N78G.pERK1_2 at time=1800 > N78G.pERK1_2 at time=3600 |
| C12 | N78G.pERK1_2 at time=300 > N78G.pERK1_2 at time=3600 |
| C13 | N78G.pERK1_2 at time=300 < T292A.pERK1_2 at time=300 |

| C14 | N78G.pERK1_2 at time=1800 < T292A.pERK1_2 at time=1800 |
|---|---|
| C15 | N78G.pERK1_2 at time=3600 < T292A.pERK1_2 at time=3600 |
| C16 | N78G.pERK1_2 at time=300 > T292D.pERK1_2 at time=300 |
| C17 | N78G.pERK1_2 at time=1800 > T292D.pERK1_2 at time=1800 |
| C18 | N78G.pERK1_2 at time=3600 > T292D.pERK1_2 at time=3600 |

**Supplemental Table 3 (N78G constraint glossary):**
This table provides a glossary for the constraint labels (C1–C18) used in Supplemental Table 6 for the N78G model. Each entry is a BPSL statement that formalizes a qualitative observation. Time points in the statements are given in seconds.

| Constraint # | T292A Model |
|---|---|
| C1 | T292A.MEK_pRDS at time=300 > T292A.MEK_pRDS at time=1800 |
| C2 | T292A.MEK_pRDS at time=300 > T292A.MEK_pRDS at time=3600 |
| C3 | T292A.MEK_pRDS at time=1800 > T292A.MEK_pRDS at time=3600 |
| C4 | T292A.MEK_pRDS at time=300 > T292D.MEK_pRDS at time=300 |
| C5 | T292A.MEK_pRDS at time=1800 > T292D.MEK_pRDS at time=1800 |
| C6 | T292A.MEK_pRDS at time=3600 > T292D.MEK_pRDS at time=3600 |
| C7 | T292A.pERK1_2 at time=300 > T292A.pERK1_2 at time=1800 |
| C8 | T292A.pERK1_2 at time=1800 > T292A.pERK1_2 at time=3600 |
| C9 | T292A.pERK1_2 at time=300 > T292A.pERK1_2 at time=3600 |
| C10 | T292A.pERK1_2 at time=300 > T292D.pERK1_2 at time=300 |
| C11 | T292A.pERK1_2 at time=1800 > T292D.pERK1_2 at time=1800 |
| C12 | T292A.pERK1_2 at time=3600 > T292D.pERK1_2 at time=3600 |

**Supplemental Table 4 (T292A constraint glossary):**
This table provides a glossary for the constraint labels (C1–C12) used in Supplemental Table 6 for the T292A model. Each entry is a BPSL statement that formalizes a qualitative observation. Time points in the statements are given in seconds.

| Constraint # | T292D Model |
|---|---|
| C1 | T292D.MEK_pRDS at time=300 > T292D.MEK_pRDS at time=1800 |
| C2 | T292D.MEK_pRDS at time=300 > T292D.MEK_pRDS at time=3600 |
| C3 | T292D.MEK_pRDS at time=1800 > T292D.MEK_pRDS at time=3600 |
| C4 | T292D.pERK1_2 at time=300 > T292D.pERK1_2 at time=1800 |
| C5 | T292D.pERK1_2 at time=1800 > T292D.pERK1_2 at time=3600 |
| C6 | T292D.pERK1_2 at time=300 > T292D.pERK1_2 at time=3600 |

**Supplemental Table 5 (T929D constraint glossary):**
This table provides a glossary for the constraint labels (C1–C30) used in Supplemental Table 6 for the T292D model. Each entry is a BPSL statement that formalizes a qualitative observation. Time points in the statements are given in seconds.

| Constraint # | WT Model | KO Model | N78G Model | T292A Model | T292D Model |
|---|---|---|---|---|---|
| C1 | 100 | 100 | 100 | 100 | 100 |
| C2 | 100 | 100 | 100 | 100 | 100 |
| C3 | 100 | 100 | 100 | 100 | 100 |
| C4 | 100 | 100 | 100 | 100 | 100 |
| C5 | 100 | 100 | 100 | 100 | 0 |
| C6 | 100 | 0.2 | 100 | 100 | 100 |
| C7 | 100 | 100 | 100 | 100 | -- |
| C8 | 100 | 100 | 100 | 100 | -- |
| C9 | 100 | 100 | 100 | 100 | -- |
| C10 | 100 | 0 | 100 | 100 | -- |
| C11 | 100 | 100 | 100 | 100 | -- |
| C12 | 100 | 100 | 100 | 100 | -- |
| C13 | 100 | 100 | 100 | -- | -- |
| C14 | 100 | 100 | 100 | -- | -- |
| C15 | 99.8 | 100 | 100 | -- | -- |
| C16 | 100 | 0 | 100 | -- | -- |
| C17 | 100 | 0 | 100 | -- | -- |
| C18 | 100 | 0 | 100 | -- | -- |
| C19 | 0 | 0 | -- | -- | -- |
| C20 | 100 | 0 | -- | -- | -- |
| C21 | 100 | 0 | -- | -- | -- |
| C22 | 0 | 100 | -- | -- | -- |
| C23 | 100 | 100 | -- | -- | -- |
| C24 | 100 | 100 | -- | -- | -- |
| C25 | 100 | -- | -- | -- | -- |
| C26 | 100 | -- | -- | -- | -- |
| C27 | 100 | -- | -- | -- | -- |
| C28 | 100 | -- | -- | -- | -- |
| C29 | 100 | -- | -- | -- | -- |
| C30 | 100 | -- | -- | -- | -- |

**Supplemental Table 6:** Percentage of accepted MCMC samples that satisfy the indicated constraint. A total of 90 constraints on predicted system behavior were defined using BPSL. Each entry is the percent of sampled parameter values yielding consistency with the indicated constraint. An entry of 100% indicates that a constraint was satisfied for sampled parameter sets, whereas an entry of 0% indicates that the constraint was never satisfied. Constraints are labeled C1–C30 and listed in the same top-down order as in each model's corresponding PROP file. Columns indicate individual models (e.g., the WT model), and rows indicate specific constraints. An entry of "--" indicates that a constraint was not applicable. See Supplemental Tables 1–5 for the BPSL statements that define the constraints.

| | Original Parameterization | PyBioNetFit Parameterization |
|---|---|---|
| Objective Funtion Score | 40.0 | 24.0 |

**Supplemental Table 7:** Overall objective function score for both Kocieniewski and Lipniacki's (2013) parameterization and PyBioNetFit's parameterization. Objective function scores were calculated using PyBioNetFit's Sum of Squares objective function and Differential Evolution fitting algorithm. Lower scores indicate better fits.

| Time (s) | Species | Experimental Data (AU) | RMSE (PyBioNetFit) | RMSE (Original) |
|---|---|---|---|---|
| **0** | pSOS1 | 0.0 | 0.0 | 0.0 |
| **0** | pEGFR | 1.5 | 1.5 | 1.5 |
| **0** | pERK | 0.9 | 0.9 | 0.9 |
| **300** | pSOS1 | 4.2 | 1.776 | 1.846 |
| **300** | pEGFR | 10.0 | 0.105 | 0.213 |
| **300** | pERK | 9.9 | 0.496 | 0.028 |
| **600** | pSOS1 | 9.5 | 0.539 | 0.39 |
| **600** | pEGFR | 9.6 | 0.277 | 1.879 |
| **600** | pERK | 9.0 | 0.245 | 0.624 |
| **900** | pSOS1 | 10.0 | 0.536 | 0.038 |
| **900** | pEGFR | 6.1 | 0.61 | 0.258 |
| **900** | pERK | 7.9 | 0.979 | 1.546 |
| **1800** | pSOS1 | 9.5 | 0.251 | 0.977 |
| **1800** | pEGFR | 2.9 | 0.836 | 0.429 |
| **1800** | pERK | 5.0 | 0.391 | 2.368 |

| **3600** | pSOS1 | 6.9 | 0.145 | 2.809 |
|---|---|---|---|---|
| **3600** | pEGFR | 1.9 | 1.579 | 1.421 |
| **3600** | pERK | 3.5 | 2.984 | 3.029 |

**Supplemental Table 8:** RMSE values for PyBioNetFit outputs vs original parameterization outputs at each time point for the data in the WT.exp file, available in the supplemental setup files in the LANL GitHub. Experimental AU data are compared to model outputs for each species (pSOS1, pEGFR, pERK), at six experimental sampling times, across both PyBioNetFit's parameterization and the original parameterization. The statistic shown is Root Mean Standard Error, highlighting the distance each model's parameterization output from the experimental data points. Lower RMSE scores is associated with a better fit to the data.

| **Parameterization** | **Global RMSE** |
|---|---|
| **PyBioNetFit** | 1.076 |
| **Original** | 1.47 |

**Supplemental Table 9:** Global RMSE scores for each model parameterization's output, considering all time points and experimental data points across all 3 species (Sos1, EGFR, ERK) as seen in Supplemental Table 8. Lower RMSE scores indicate a better fit to the experimental data.

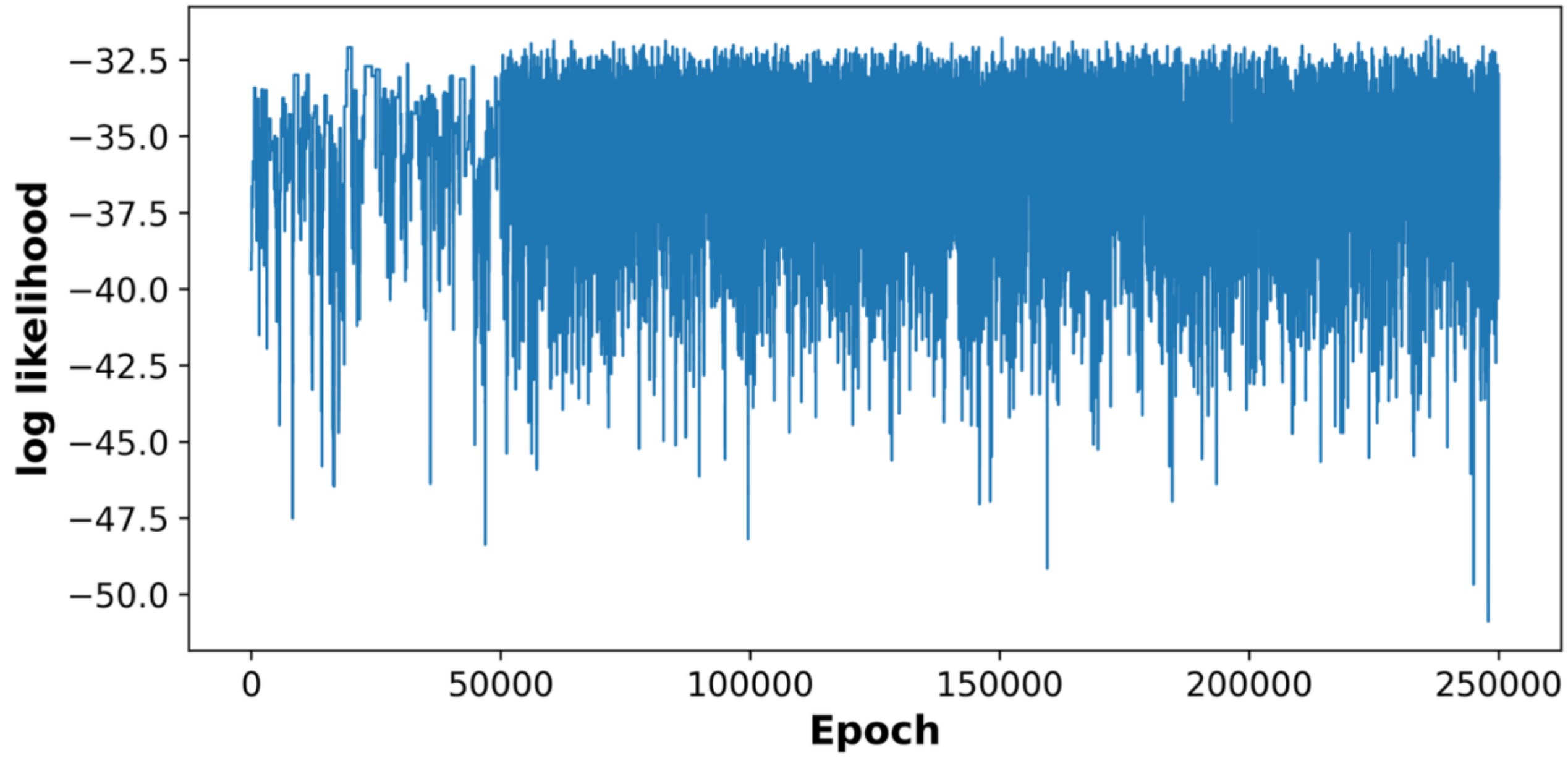

**Supplementary Figure 1:** Trace plot of log-likelihood values across 250,000 production iterations of MCMC sampling.

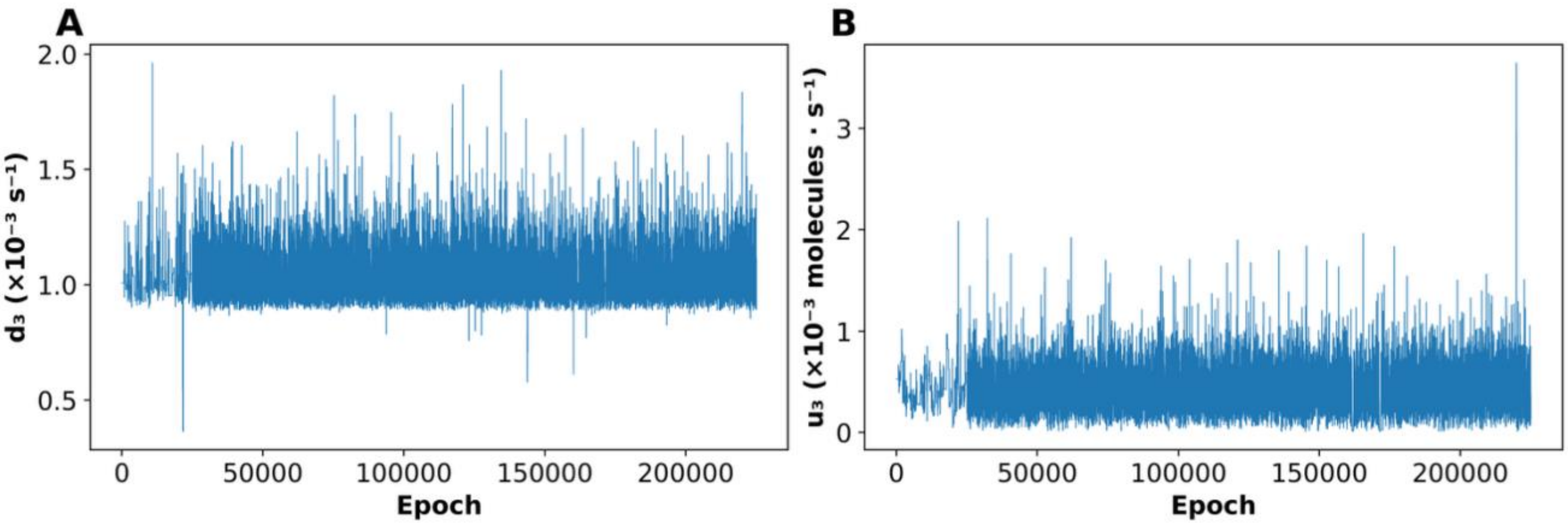

**Supplementary Figure 2:** Trace plots for each of two parameters across 250,000 iterations of production sampling. Panel A shows the trace plot for $d_3$, and panel B shows the trace plot for $u_3$.

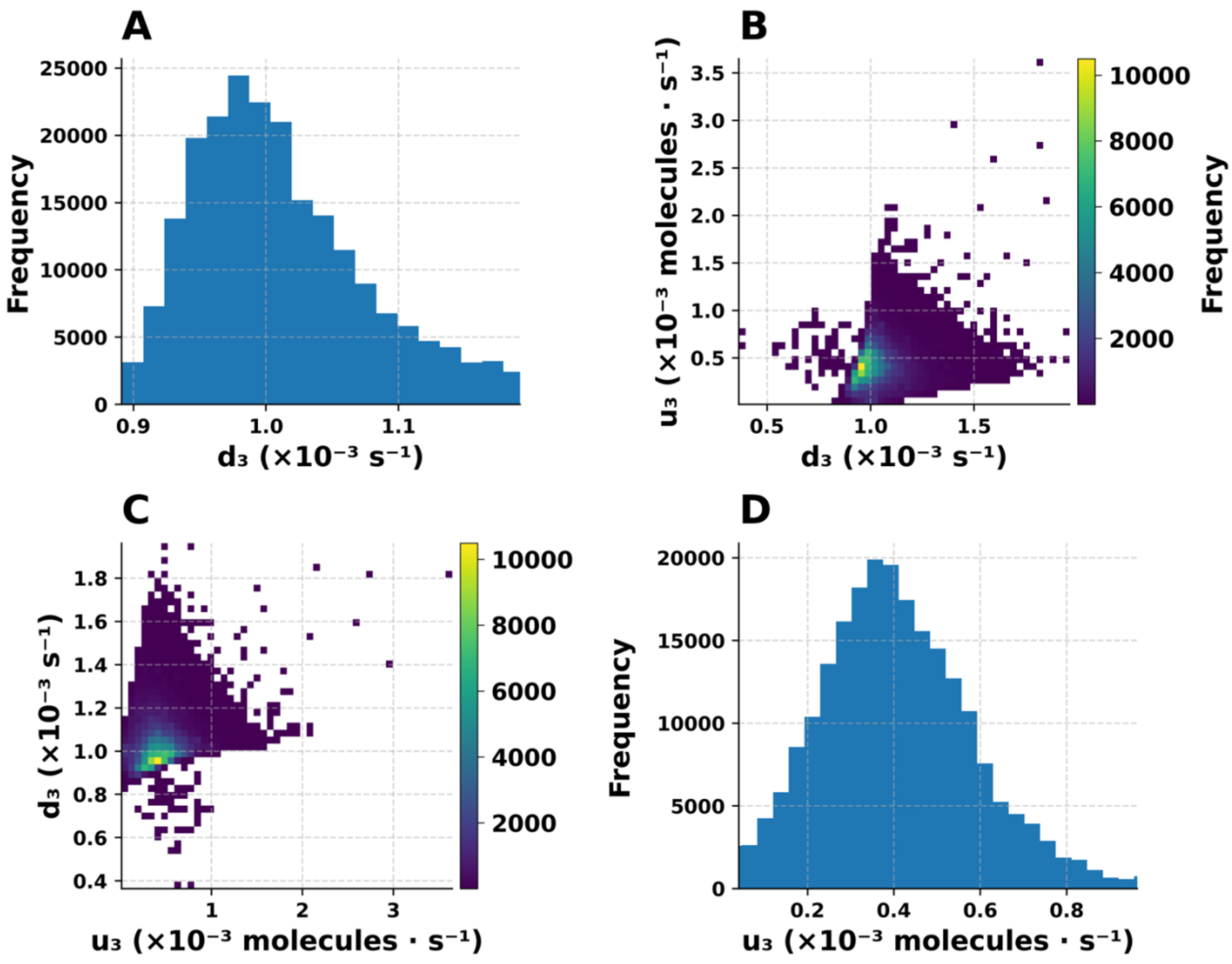

**Supplementary Figure 3:** Pairs plots of posterior samples for the rate constants $d_3$ and $u_3$. (A) Marginal posterior for $d_3$. (D) Marginal posterior for $u_3$. (B, C) Joint posterior density, i.e., the continuous two-dimensional density estimate of the posterior over $(d_3, u_3)$. Colors indicate relative sample density, from white (lowest) to yellow (highest).

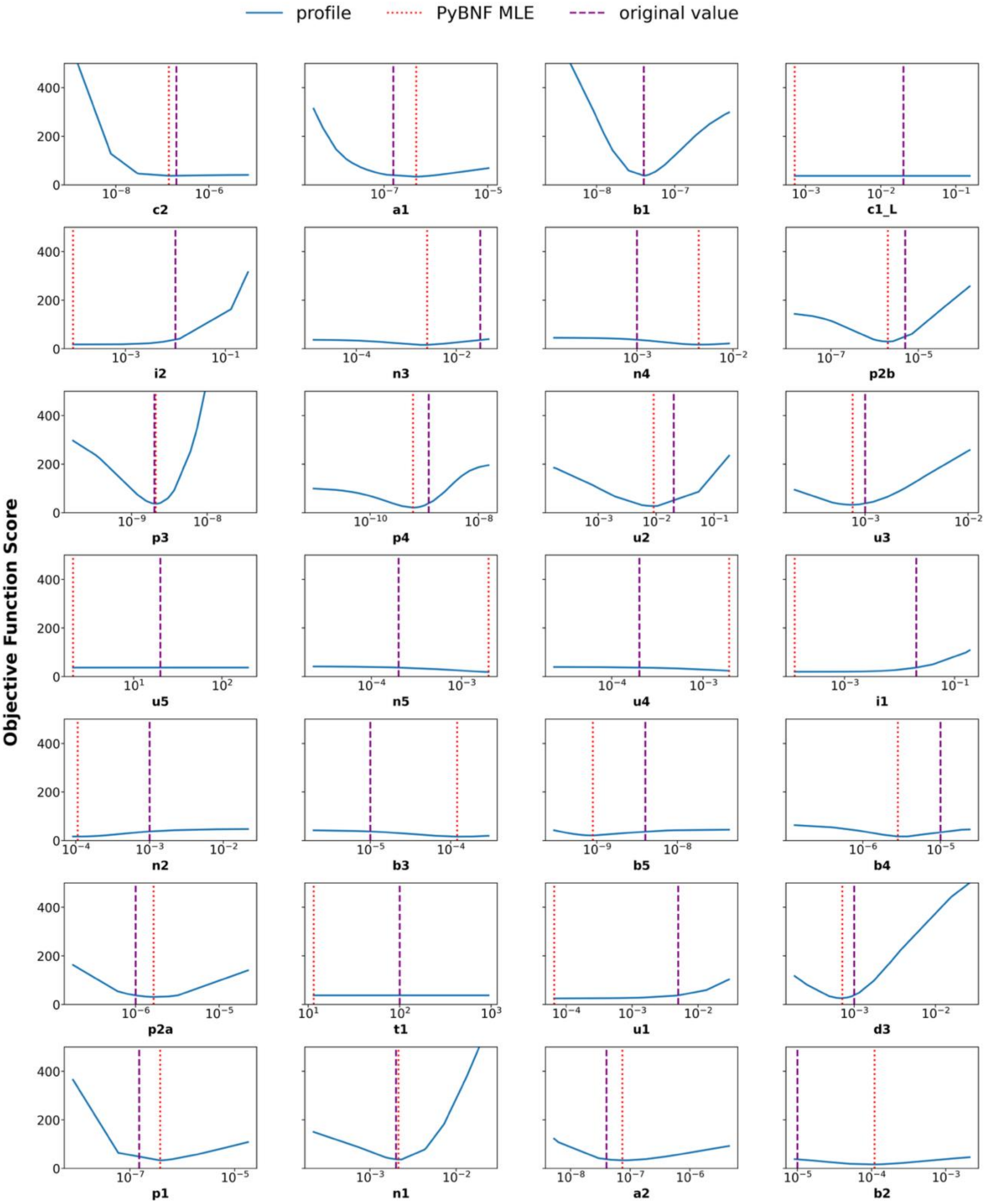

**Supplementary Figure 4:** A profile likelihood table of all 28 parameters used for parameterization with PyBioNetFit. The x axis (log scale) for each plot shows what parameter value corresponds to which objective function score on the y axis. Vertical, red-dotted lines indicate the maximum likelihood estimate for that parameter. Vertical, purple-dotted lines indicate the original parameter

value used by Kocieniewski and Lipniacki (2013). Parameters that are identifiable should have curved, bell-shaped, lines (blue lines) that decrease in objective function score until they reach their maximum likelihood estimates (red-dotted lines). Parameters that are not identifiable appear flat or near flat.

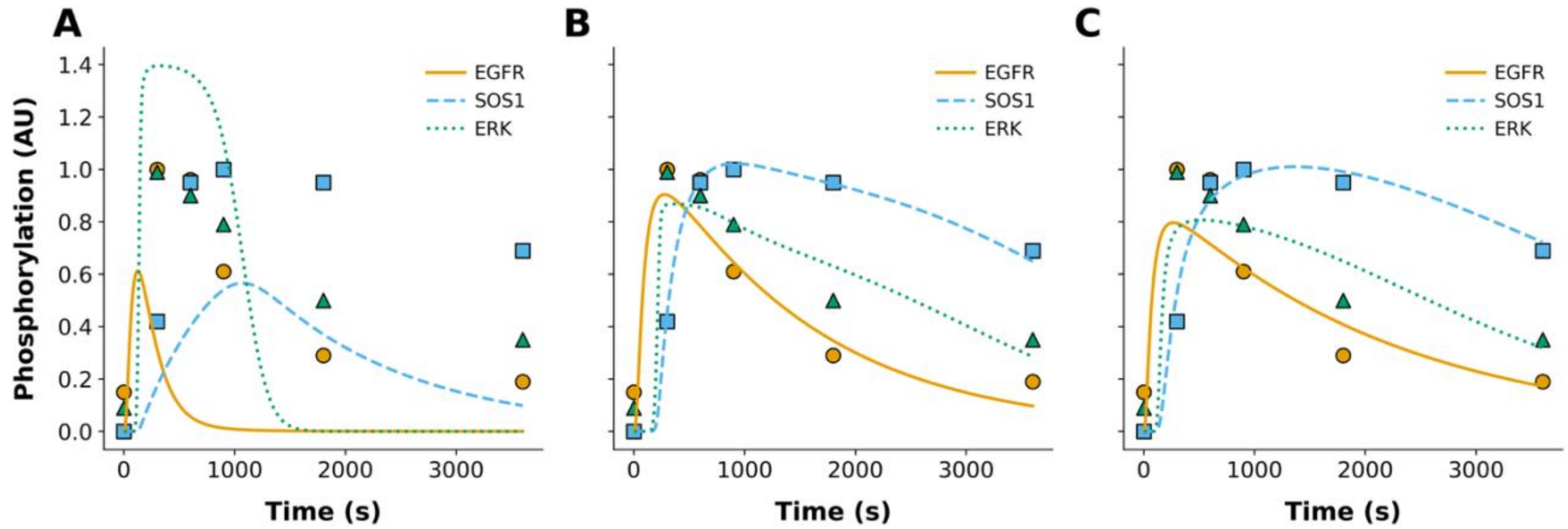

**Supplementary Figure 5:** Ablation analysis to assess the contributions of only BPSL constraints, only experimental data, and a combination of both data types to the overall fit to the experimental data. Panel A shows PyBioNetFit's overall fit to experimental data only using only BPSL statements (qualitative data) with an objective function value of **39**. Panel B shows PyBioNetFit's overall fit to the experimental data using only experimental data, with an objective function score of **8**. Panel C shows PyBioNetFit's overall fit to the experimental data using both BPSL statements (qualitative constraints) and experimental data with an objective function score of **24**. In all three instances, PyBioNetFit's differential evolution (DE) fitting algorithm was used along with PyBioNetFit's sum of squares (SOS) objective function to evaluate "closeness of fit". Lower objective function values indicate a closer fit between the projected time courses and the experimental data. Because there is experimental data available only to the Wild-Type model (WT), we could only use the WT-specific constraints (BPSL statements) to make a fair comparison between each plot. Thus, there were **6** WT BPSL statements (qualitative data) used in this ablation analysis. Using only BPSL statements (Panel A) PyBioNetFit's parameterization followed **6/6 constraints**. Using only experimental data (Panel B), PyBioNetFit's parameterization followed **6/6 constraints**. Using both BPSL statements and experimental data (Panel C), PyBioNetFit's parameterization followed **6/6 constraints**. These results demonstrate that incorporating both quantitative and qualitative constraints yields the most balanced and accurate parameterization, improving reproducibility without sacrificing fit quality.